\documentclass{JINST}

\usepackage{units}		
\usepackage{comment}	

\title{Studies of helium poisoning of a Hamamatsu R5900-00-M16 photomultiplier}

\author{	R.~Ospanov$^a$, 
		M. Kordosky$^a$,
		K. Lang$^a$\thanks{Corresponding author.}~ 
		J. Liu$^a$,
		T. Osiecki$^a$,
		M. Proga$^a$,
		and P. Vahle$^a$\\
\llap{$^a$}The University of Texas at Austin,
		 Department of Physics C1600,
		 1 University Station, Austin, TX 78712-0264, USA\\
  E-mail: \email{lang@physics.utexas.edu}}

\abstract{We report results from studies of the helium poisoning of a 16-anode
photomultiplier tube R5900-00-M16 manufactured by Hamamatsu Photonics.
A tube was immersed in pure helium for a period of about four months
and was periodically monitored using a digital oscilloscope.
Our results are based on the analysis of waveforms triggered by 
the dark noise pulses. Collected data yield evidence of after-pulses
due to helium contamination of the tube. The probability of after-pulsing 
increased linearly with the exposure time to helium but the phototube suffered 
only a small drop in gain, indicating generally strong resilience to helium 
poisoning.
}

\keywords{helium poisoning; photomultipliers; PMT; after-pulsing; multi-anode photomultipliers}

\begin{document}


\section{Introduction}

We report results from studies of helium poisoning of a 16-anode 
photomultiplier tube (PMT), model R5900-00-M16 (M16), manufactured 
by Hamamatsu Photonics. This work was motivated by the need to
better evaluate the lifetime of the MINOS Far 
detector~\cite{MINOS_Proposal,MINOS_TDR,Lang:2001rw}
which employed about 1,600 of these PMTs~\cite{M16-1600}.

The MINOS Far detector was in continuous operation since its
completion in the Summer of 2003 until Summer 2016~\cite{1st-MINOS-PRL}.
Initially, the detector recorded cosmic ray data, and since 
the beginning of 2005, MINOS commenced its beam operations using 
the NuMI neutrino beam at Fermilab. 

The MINOS Far detector~\cite{Lang:2001rw} was installed 
in a cavern of the Soudan Underground Laboratory 
located about \unit[705]{m} underground.  
This shielding  of 2,070 meter-water-equivalent
from cosmic rays results in low muon flux 
of about \unit[0.5]{Hz} in the entire \unit[5.4]{kton} detector. 
The rates in the M16 PMTs are dominated by 
natural radioactivity (gammas) interacting in the scintillator and 
the spontaneous emission of light from wavelength-shifting 
fibers, as reported earlier~\cite{wls-fiber}. 
Because of these relatively low 
rates of about \unit[1-2]{kHz} per PMT, mainly due to single
photoelectron pulses, and stable environmental conditions 
underground, we had expected a long mean lifetime for the M16 
photomultipliers employed by MINOS.

However, a neighboring experimental hall in the Soudan 
laboratory housed the CDMS-II experiment which 
searched for cold dark matter and which employed 
a cryogenic helium system~\cite{CDMS-2}.
During routine operations of CDMS-II, the concentration of helium in 
the MINOS cavern was about \unit[60]{ppm}, but jumped to about 
\unit[2000]{ppm} during initial liquid helium transfers in 
the CDMS-II hall.  This can be compared to the concentration of helium 
in the earth's atmosphere near the earth's surface which in detail
depends on location but generally is at the level of about \unit[5-6]{ppm}.
As it is well established, in addition to high rate or current operations,
the principal factors affecting the aging  of most of 
photomultiplier tubes is the helium poisoning (i.e., contamination
of the PMT vacuum by helium atoms which can permeate from
the atmosphere).
Though CDMS-II installed later its own venting system for
helium handling, the proximity of large amounts 
of this gas elevated our concerns  regarding the lifetime 
of the MINOS PMTs and directly inspired the tests reported here. 

Performance of a PMT depends, among other factors,
 on the quality of vacuum inside 
the tube. Residual molecules or atoms trapped inside a tube  
may be ionized in collisions with (photo)electrons. Such
scatterings decrease the electron energy and the effective
collection efficiency of the first dynode of a PMT. 
Positive ions resulting from these collisions can 
be accelerated toward the cathode where their 
impact may lead to secondary electron emission and the onset
of an additional and time-correlated dynode amplification 
chain, observed as after-pulsing. 
The probability of the ionization of the residual gas depends 
primarily on the gas density, the electron energy, 
and the distance which photoelectrons travel between a photocathode 
and the first dynode, or between the dynodes. 
These processes, generally, decrease an average 
PMT gain and increase the noise due to after-pulses. 

In our studies, we immersed an M16 PMT in helium for about four
months, and periodically monitored the PMT dark noise using
a digital oscilloscope. We analyzed the time and amplitude 
structures of recorded waveforms.  Our results include 
studies of the change of the PMT gain and the  after-pulsing
probability as a function of the cumulative exposure time 
to helium. 

In sections 2 and 3  we describe the experimental 
setup and present our main findings. In section 4 we provide simple
estimates for  probability and timing of after-pulses in M16 PMTs.
In section 5 we briefly discuss our results.

\section{Experimental setup and data sets}

\begin{figure}[h]
\begin{center}
\includegraphics[width=.45\textwidth]{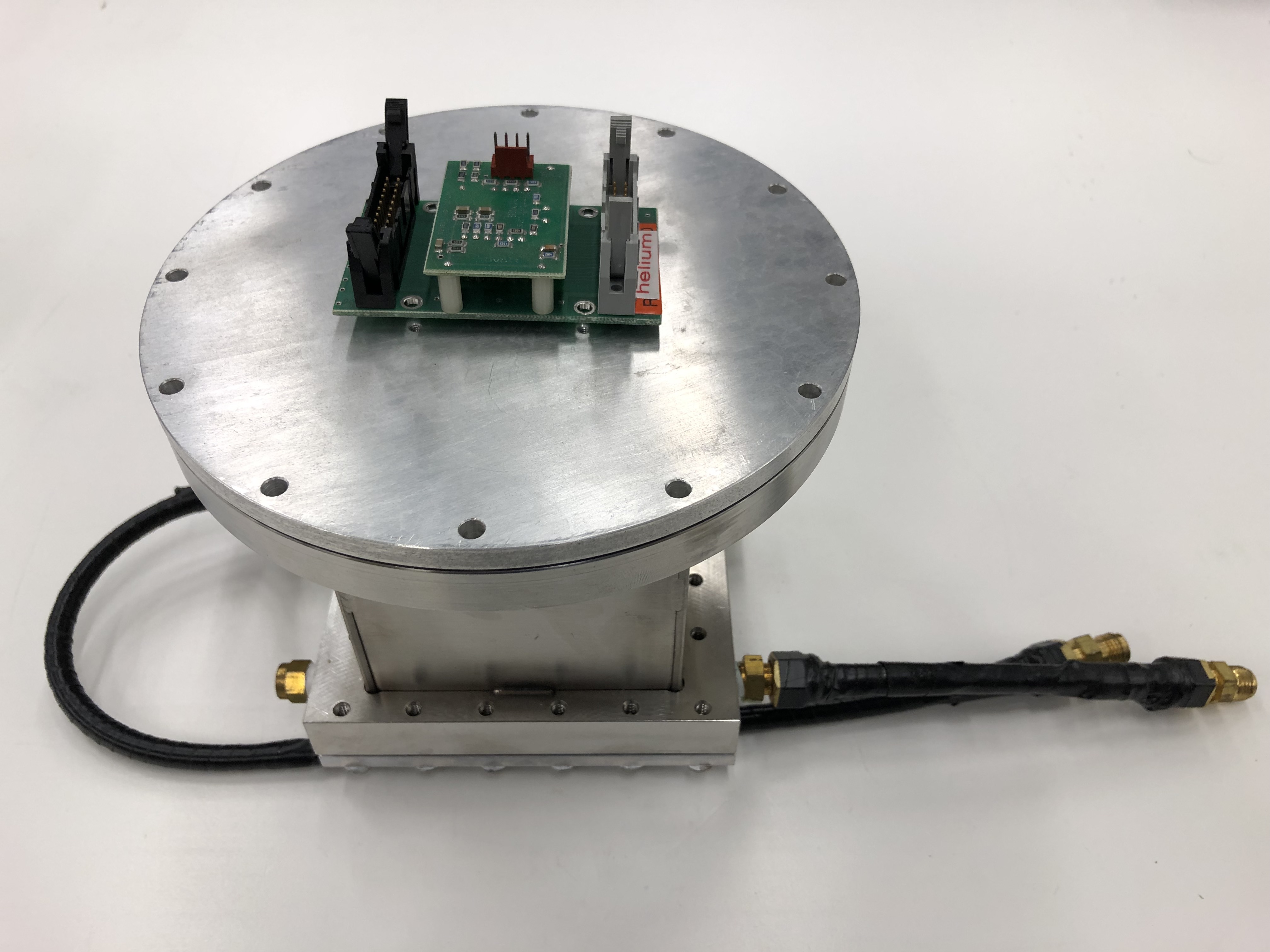}
\hskip0.25in
\includegraphics[width=.45\textwidth]{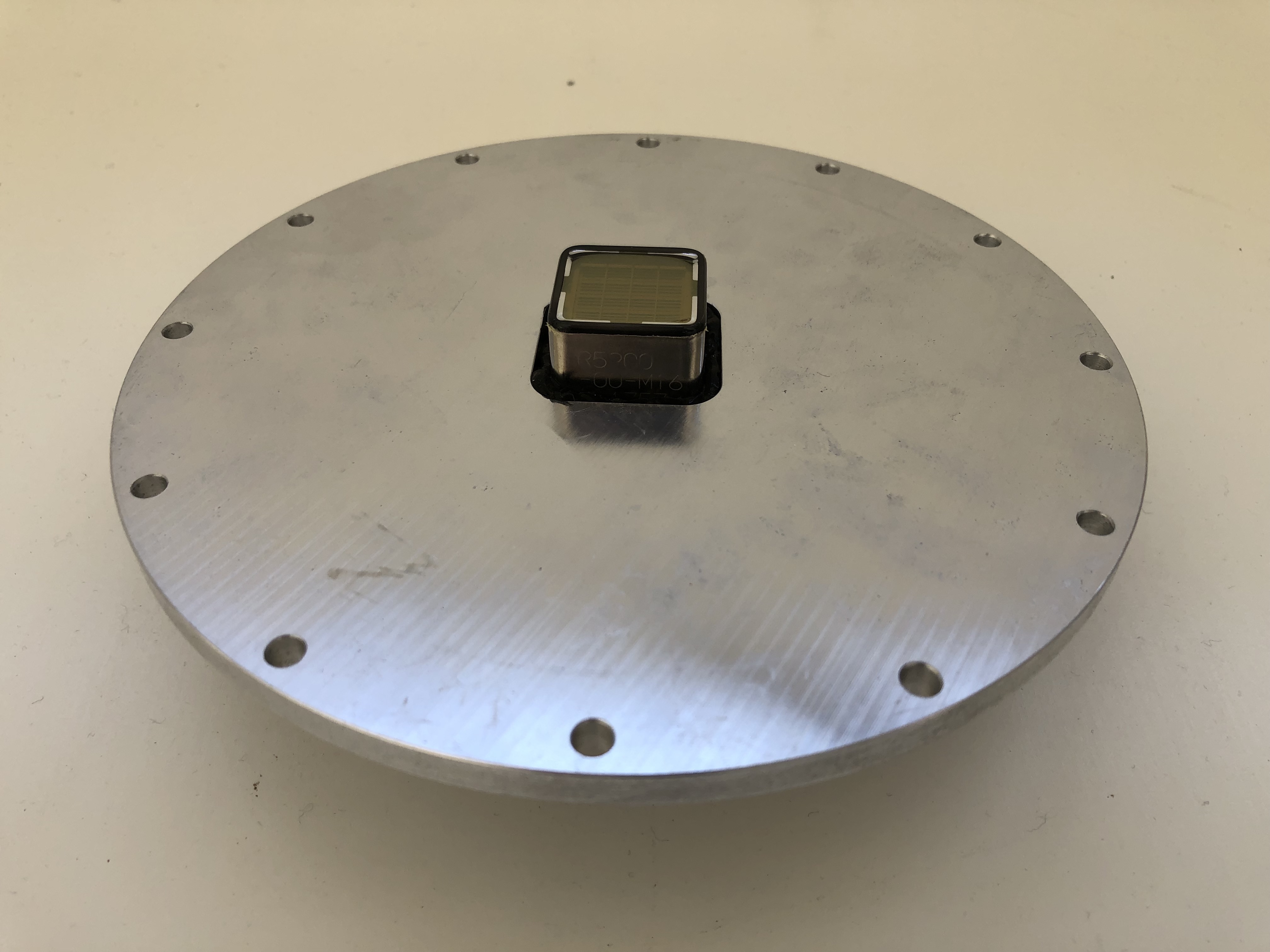}
\caption{The main part of the experiment -- the helium vessel --  is shown on the left.
The photomultiplier faced helium through a sealed opening in the top flange of vessel.}
\label{fig:M16-helium-setup}
\end{center}
\end{figure}

The experimental setup, shown in Figure~\ref{fig:M16-helium-setup}, comprised five main components: 
an M16 PMT with its custom-designed printed-circuit base, a steel container enclosing
the helium and the PMT, a dark box, a digital oscilloscope, 
and a desktop computer which acquired data from the
oscilloscope using LabView~\cite{LabView}. 

\begin{figure}[b]
\begin{center}
\includegraphics[bb=700 0 1770 1850,width=.39\textwidth]
{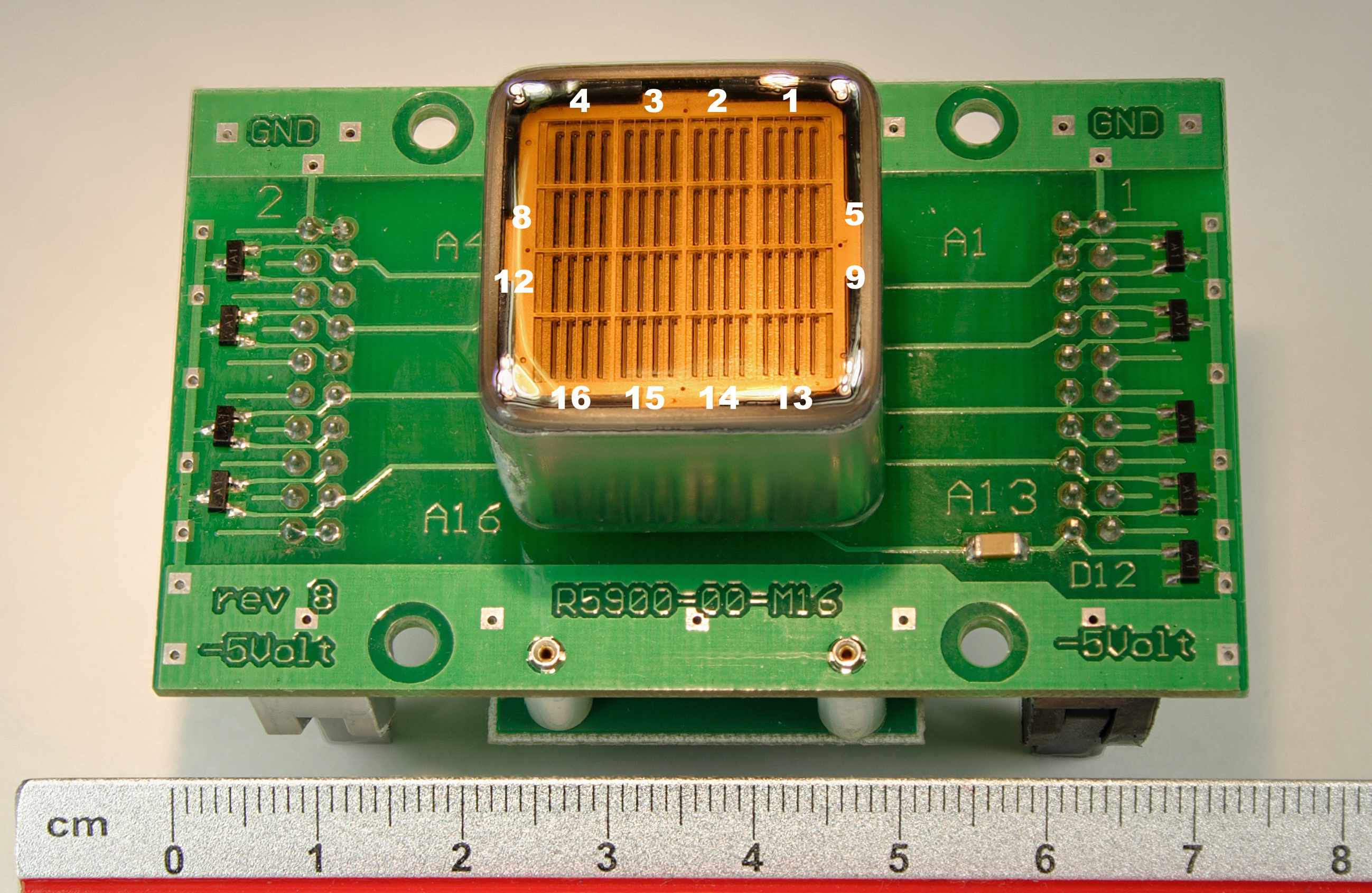}
\caption{A picture of a M16 PMT with a base. The tube has dimensions 
of \unit[24]{mm} $\times$ \unit[24]{mm} $\times$ \unit[20]{mm} and
the casing is made out of a metal alloy called KOVAR. 
Helium can permeate into the tube
primarily through the photocathode window \unit{0.8}{mm}-thick,
made out of borosilicate glass and the epoxy sealing pins for dynodes. 
In the picture we marked pixel
numbers in accordance with the factory's numbering scheme 
which was also used in MINOS.}
\label{fig:M16}
\end{center}
\end{figure}

M16 PMT has a compact, almost cubical shape of dimensions 
\unit[24]{mm} $\times$ \unit[24]{mm} $\times$ \unit[20]{mm}.  
Figure~\ref{fig:M16} shows a close-up photograph of such 
a PMT with an attached base, which serves
as a dynode voltage divider and an anode signals router~\cite{M16-1600}.
The square front window is \unit{0.8}{mm}-thick and made out 
of borosilicate glass.  All other sides of the M16 casing 
are made of KOVAR, a metal alloy
with a small coefficient of thermal expansion~\cite{Hamamatsu-private}.
The M16 dynode and anode pins protrude from the back of 
the tube through insulated epoxy-sealed holes.

For our tests we selected a tube with a low dark noise rate and 
a high gain of about $2 \times 10^7$, achieved at negative \unit[1000]{V},
the maximum photocathode potential recommended by the manufacturer. 
At such a high gain,
pulses due to single photoelectrons are easily detected by 
a digital oscilloscope~\cite{LeCroy}. 
The insulated PMT (the KOVAR casing is at the same
potential as the photocathode) was installed facing inward 
 and sealed in the lid of a small steel container. 
The back (i.e., pin) side of the tube with a mounted base was
outside of the helium atmosphere.
The steel container had an inlet and an outlet 
for the helium flow and was placed in a dark box. 
Special care was taken to assure that all gas 
and signal connections did not compromise the light-tightness
of the dark box. Pure helium was flowed at \unit[10-20]{cc/min} 
and resulted in about 100 volume exchanges per day.
Due to the gas flow, the pressure of helium in the container 
was only slightly above the atmospheric pressure. 

The compact architecture of the M16 PMT, thus short
electron and ion drift distances, led us to
expect relatively short (\unit[10-100]{ns}) time 
separation between dark noise pulses and their 
helium after-pulses. This guided us to study waveforms 
triggered by single photoelectrons due to the thermionic 
(dark) noise from the photocathode. 
We decided against the use of an external
light source since most light sources produce pulses lasting
tens of nanoseconds which would obscure or conceal 
a significant part of the after-pulsing. 

We connected individual anodes directly to a digital oscilloscope
which was controlled by a specially designed LabView Virtual Instrument 
application  running on MS Windows 2000. 
The oscilloscope was configured in a self-triggered
mode with a threshold set to \unit[-12]{mV}
and a maximum sampling rate of \unit{4}{GS/s}. 
This allowed the readout of waveforms with a \unit[500]{ns}-wide 
time window with the trigger time positioned at 20\% of
the scale (i.e., at \unit[100]{ns}). It took about \unit[0.2-0.3]{s} 
to acquire a single waveform. The internal dead-time of the oscilloscope 
was about \unit[10]{ms}. The dark rates of individual anodes varied 
from 3 to \unit{13}{Hz}. 

We conducted the tests over a period of about four months
(141 days)
in the course of which the tube was continuously immersed in helium.
During  the first two weeks of the test, the data (individual
pixel waveforms) were taken every day.
After this initial period, the waveforms were collected 
once every 10-20 days.  
The data acquisition took \unit[3-4]{hrs} per channel.

Each data acquisition session resulted in
80,000 waveforms  recorded for pixels 4, 5, 9, 13, and 16 
(for some checks we also acquired 150,000 waveforms 
for pixels 13 and 16).  
The total number of acquired waveforms was constrained by 
the acquisition time and available disk space. During our tests,
we monitored the high voltage (set to \unit{-1000}{V}),
and checked the temperature in the laboratory, which
stayed constant at $21 \pm 1 ^\circ$C, except toward the
end of the testing period when the temperature was a couple
of degrees lower due to more frequent automatic 
air-conditioning.

\section{Main results}

\begin{figure}[b]
\begin{center}
\includegraphics[
width=.49\textwidth]
{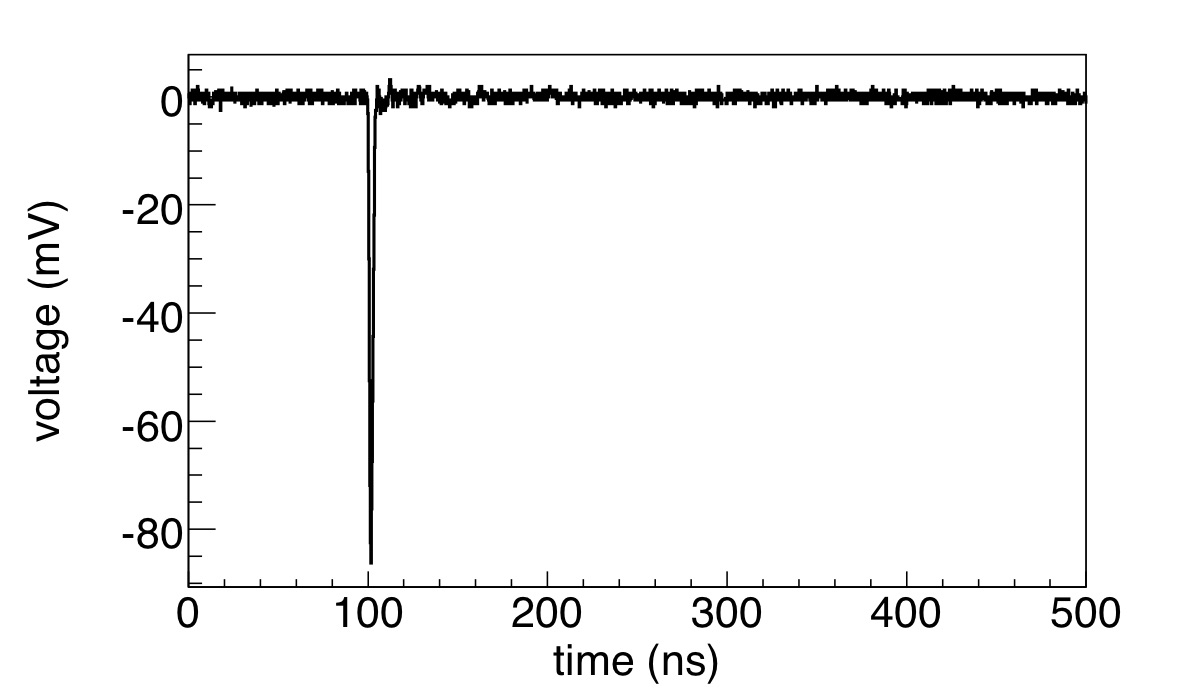}
\includegraphics[
width=.49\textwidth]
{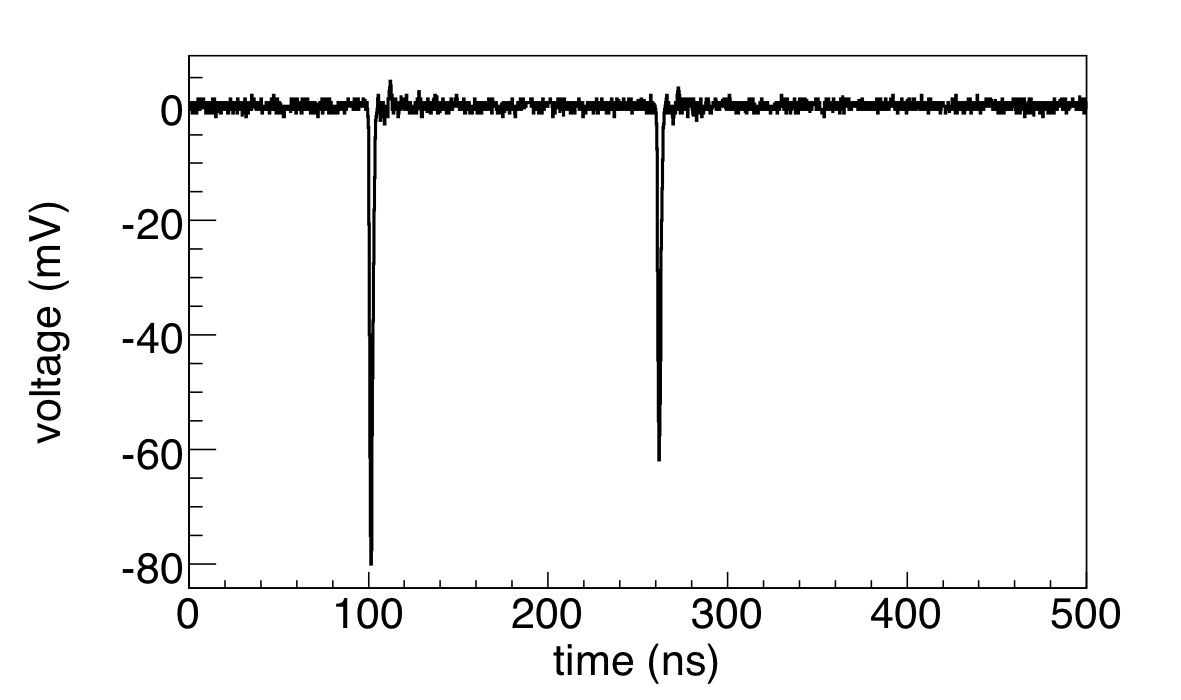}
\includegraphics[
width=.49\textwidth]
{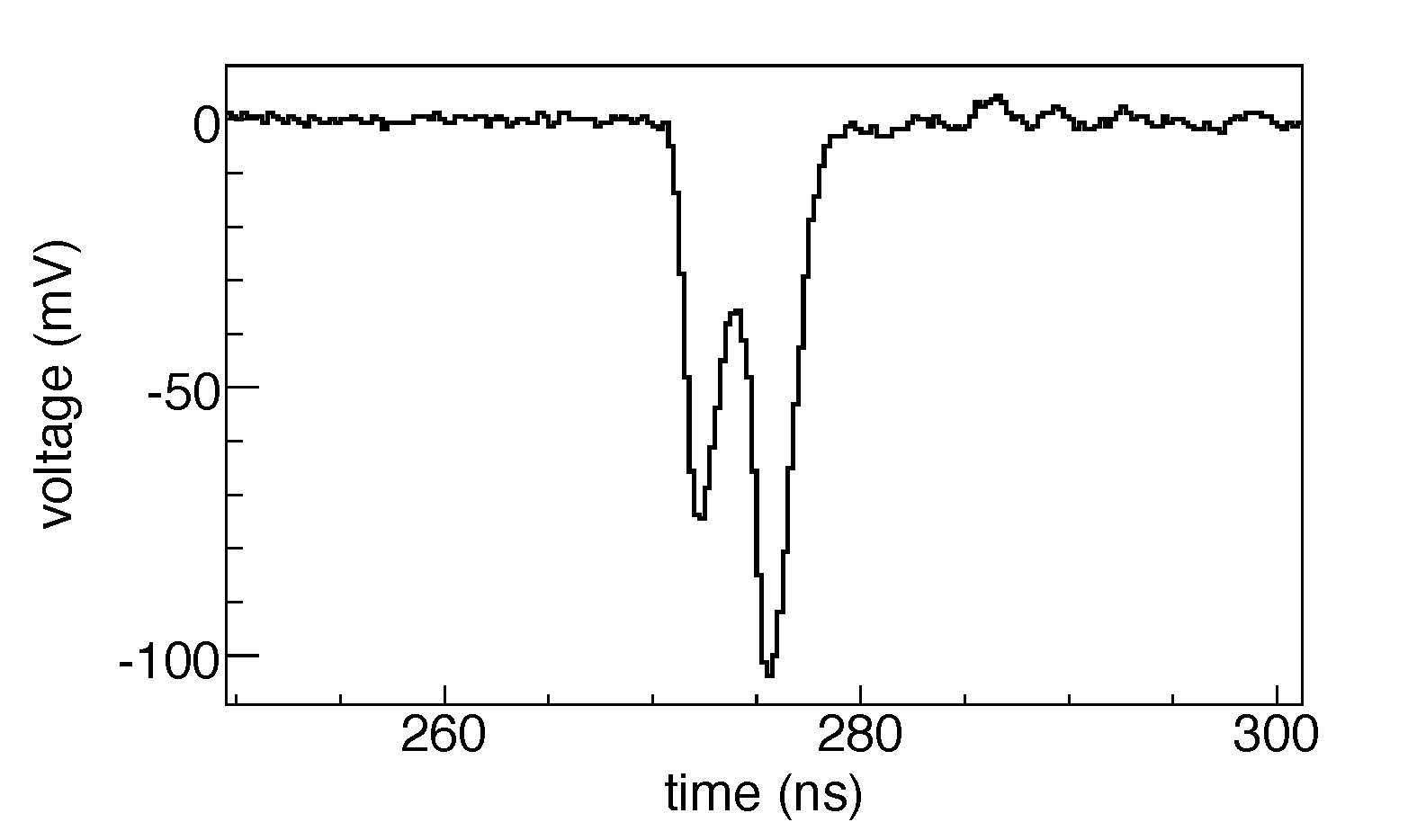}
\includegraphics[
width=.49\textwidth]
{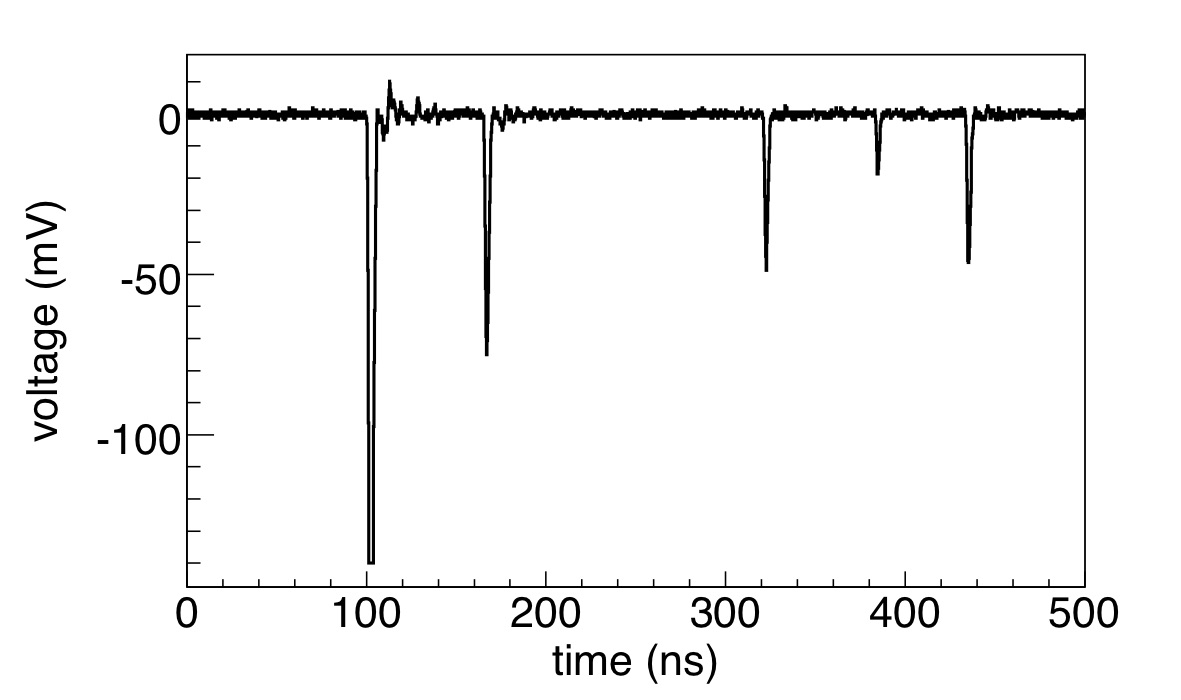}
\caption{Examples of various waveforms. 
Top-left: A typical waveform with one peak due to a single photoelectron.
Bottom-left: An example of two overlapping single photoelectron pulses
(note the difference in the time scale of this plot).
Top-right: A typical waveform with an after-pulse.
Bottom-right: A rare waveform with five after-pulses 
($\ll$ 1\% of all waveforms were of this type).}
\label{fig:waveforms}
\end{center}
\end{figure}

The main objective of the data analysis was to characterize
the time and amplitude structure of each acquired waveform.
Figure~\ref{fig:waveforms} shows four examples of waveforms 
in our data.
A single waveform comprises 2000 time bins spanning \unit[500]{ns} 
(i.e., individual bins in a waveform are \unit[250]{ps} long)
with a voltage value for each bin.
Each waveform is time and date-stamped.    
The time resolution of the oscilloscope allowed to resolve
near or overlapping pulses, as it is illustrated in the bottom-left
panel of Figure~\ref{fig:waveforms}.
The majority of the pulses have a width of about \unit[4-8]{ns}. 
A typical waveform with an after-pulse has a second pulse, as shown 
for example in 
the top-right panel of Figure~\ref{fig:waveforms}. We have also
observed more complex waveform structures,
as shown in the bottom-right panel of Figure~\ref{fig:waveforms}.

We applied a straightforward procedure to locate and measure
characteristics of pulses in each waveform. In our algorithm,
a pulse in a waveform was identified by a monotonic voltage 
change as a function of bin number denoting time.
The time bin with the maximum value of the voltage 
(i.e., the pulse amplitude) marks the time coordinate 
(position) of the pulse. At the peak amplitude, 
the sign of the voltage gradient reverses. 
Occasionally, the oscilloscope registered waveforms triggered 
by RF noise. Those were identified as waveforms with 
a large number of peaks due to sinusoidal RF pulses.
We validated the
performance of the pulse reconstruction algorithm by 
visual scanning a significant portion of data.

To minimize the effect of noise, we decided to express 
the magnitude of the pulse amplitude as the pulse charge,
$Q$, calculated as the area under the pulse:
$Q=\frac{1}{R}\int_{t_1}^{t_2}V~dt$, where $R = 50~\Omega$.
The time-limits of this integral (summation in practice) 
are defined by points (bins)
close to the voltage baseline (ground) for which the pulse
amplitude gradient changes sign or becomes zero.

\subsection{Probability of after-pulsing of the M16 }

\begin{figure}[b]
\begin{center}
\includegraphics[width=.49\textwidth]
{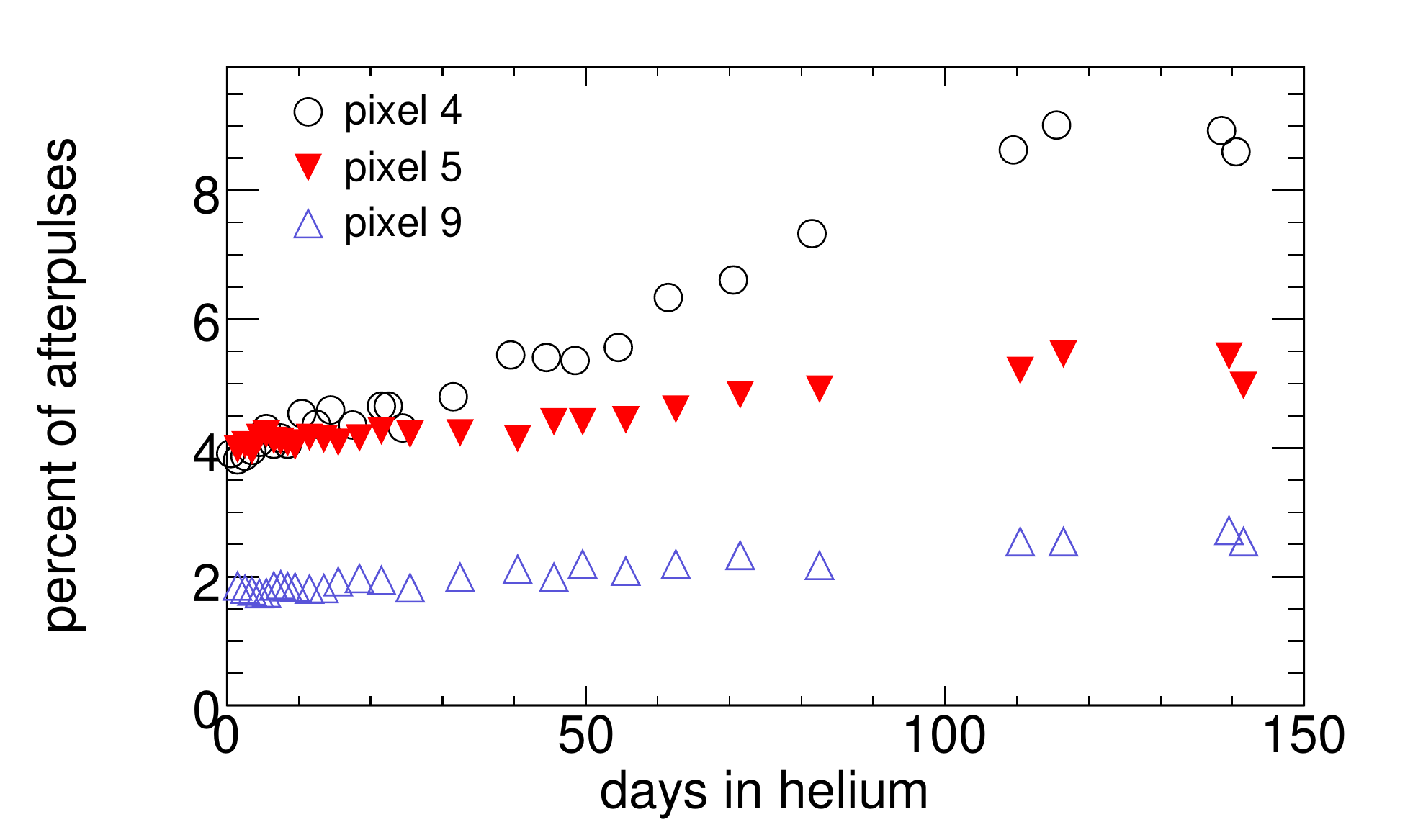}
\includegraphics[width=.49\textwidth]
{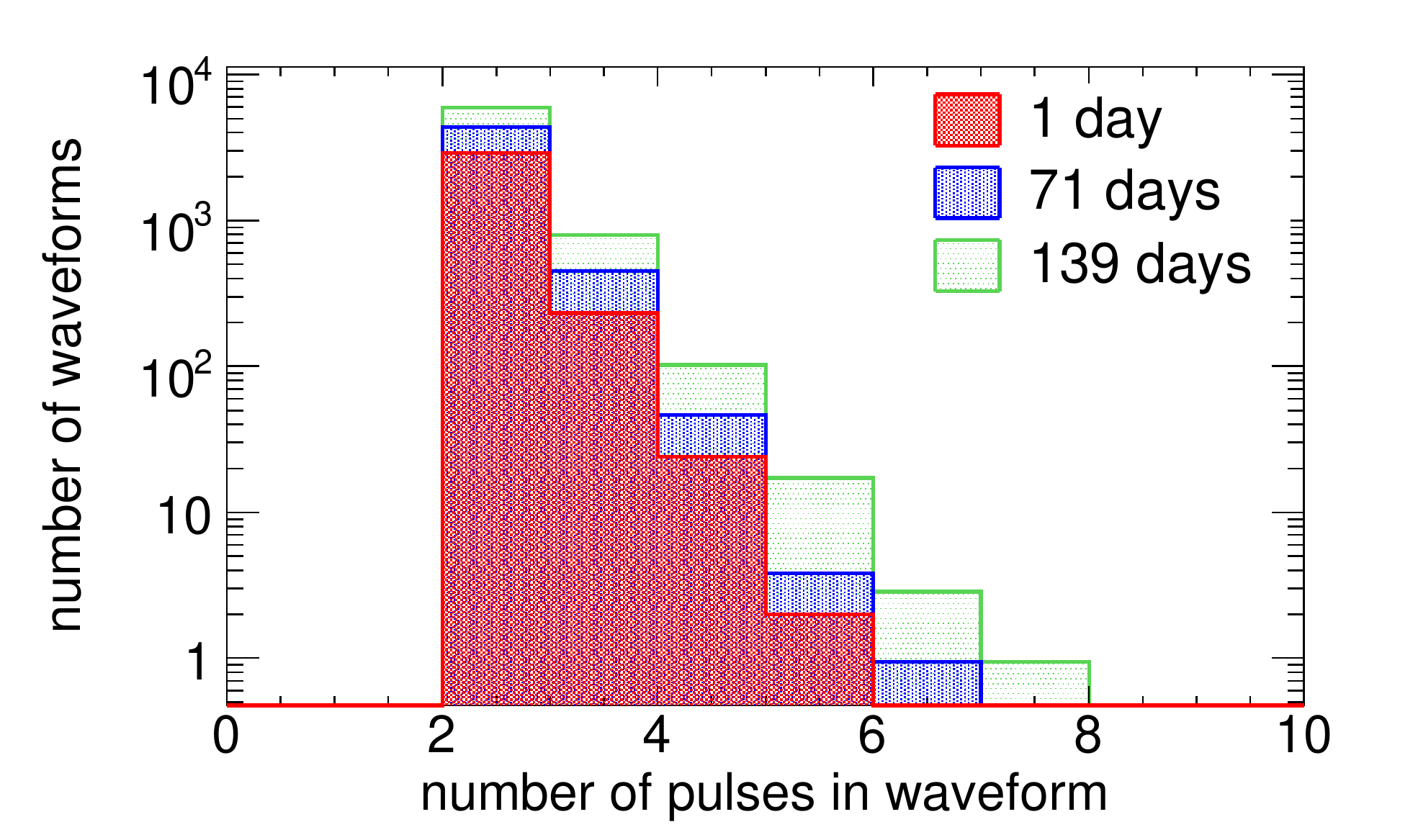}
\caption{Probability of after-pulsing.
Left: A fraction (in percent) of waveforms with two pulses as 
a function of the exposure time to helium.
Right: A distribution of fractions (in percent) of waveforms 
with more than two pulses. 
The three superposed distributions were obtained using 
data taken about two months apart (i.e., at the beginning, 
in the middle, and at the end of our tests).}
\label{fig:fraction-of-2-pulses}
\end{center}
\end{figure}

The probability of after-pulsing was taken as a fraction of waveforms 
with multiple pulses for each data set, after 
removing waveforms with the RF noise, as discussed above.
In Figure~\ref{fig:fraction-of-2-pulses} we plot the fraction of 
waveforms with two pulses as a function of the exposure time to helium. 
Pixels 4, 13 and 16 in the M16 PMT are all corner pixels. 
They are slightly larger in area and are more sensitive
to a fringe electric field.
Pixels 5 and 9 are neighboring edge pixels. 
Although we observe a clear growth of after-pulsing with time, 
both the relative and the absolute increase of after-pulsing
with time depends on the pixel. 
These differences between pixels are not well understood. 
We found no correlations between dynamic characteristics 
(like a gain and a dark noise rate) of individual  pixels
with the trends displayed in 
Figure~\ref{fig:fraction-of-2-pulses}.

As Figure~\ref{fig:fraction-of-2-pulses} indicates, the after-pulsing
was present at the start of our tests at the level of 2-4\%  
and must be attributed to the initial contamination of the PMT vacuum.
The probability of after-pulsing grew with the exposure time to
helium. For a corner pixel 4, the increase was about 4\%,
and for the edge pixels 5 and 9 the increase was about 1\% .

\subsection{Gains and widths of single pulses}

\begin{figure}[h]
\begin{center}
\includegraphics[width=.49\textwidth]{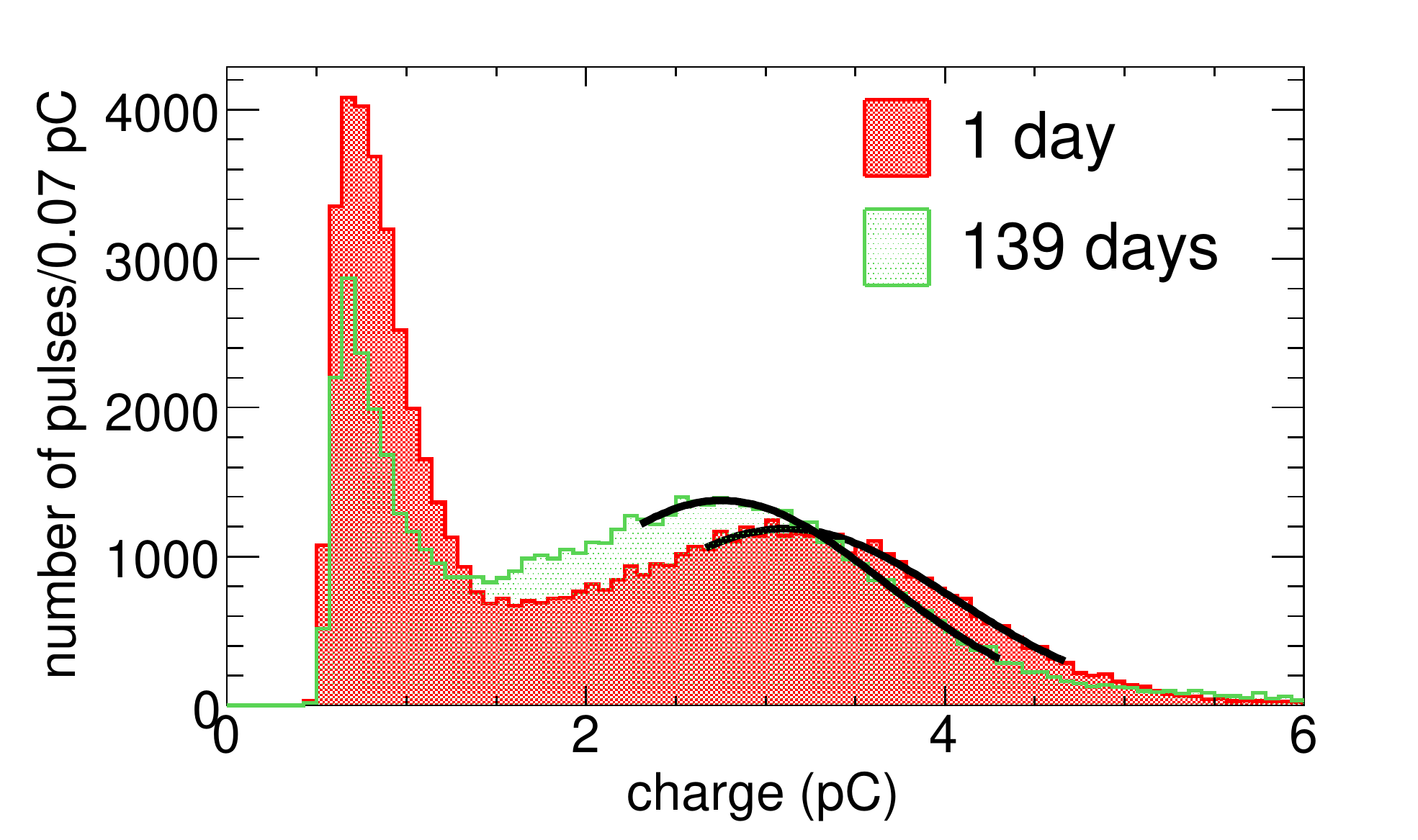}
\includegraphics[width=.49\textwidth]{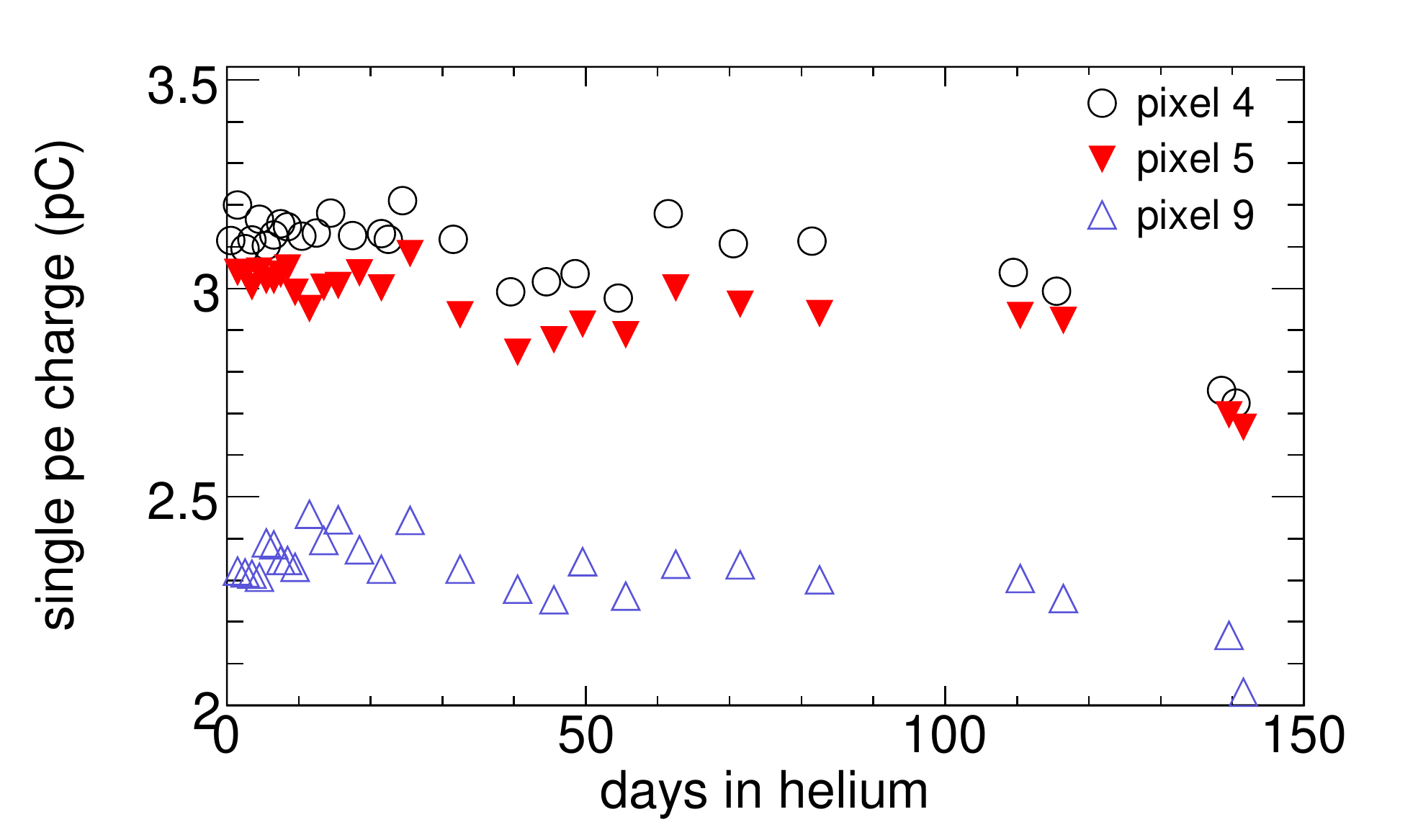}
\caption{Gain change as a function of the exposure time
to helium. Left:
Distributions of the integrated charge 
(in picocoulombs) of a single photoelectron pulse for
waveforms with only one pulse (i.e., without after-pulsing).
The two histograms are for the data taken about four months
apart, at the beginning and at the end of our tests.  
The position of the peak was initially at $3.12\pm0.02$~pC,
and it moved to $2.73\pm0.02$~pC after 139 days in helium.
Right: The mean integrated charge of a single photoelectron 
pulse, determined by fitting histograms as illustrated 
in the top plot, performed on each of the 27 data sets 
collected over four months.}
\label{fig:pix-4-gain-vs-time}
\end{center}
\end{figure}

We analyzed the gains and widths of pulses in single-pulse
waveforms (i.e., waveforms with no visible after-pulsing).
The PMT gain was calculated from the integrated charge
of single photoelectron pulses. The gain was examined  
as a function of the time exposure to helium.
The top panel of Figure~\ref{fig:pix-4-gain-vs-time} shows two 
superposed distributions of the integrated charge for waveforms
with a single pulse. The two histograms are for data taken about 
four months apart (i.e., at the beginning and at the end 
of our tests). All tested pixels yielded distributions with very 
similar features, thus,
here and in other plots we present results for pixel \#4,
unless otherwise stated.

The sharp cut-off on the left side of both distributions 
in Figure~\ref{fig:pix-4-gain-vs-time}
reflects the setting of the trigger threshold on the oscilloscope.
The left (sharp) peak is due to those photolectrons which 
suffered only partial amplification through the dynode chain, 
typically missing the first dynode, as we identified 
in our earlier studies~\cite{M16-1600}. The broad peak, 
centered around the charge value $Q \approx 3$~pC, is due to 
a single photoelectron. The position of this peak  
measures directly the gain of the pixel. The value $G$ of 
the gain is simply given by  
$ G = Q/e \approx 3\times 10^{-12}~[{\rm C}] / 
1.602\times 10^{-19}~[{\rm C}]  \approx 1.9 \times 10^{7}$.

In the bottom panel of Figure~\ref{fig:pix-4-gain-vs-time}, 
we display the mean charge of a single photoelectron 
pulse as a function of exposure time to helium. 
The value of each mean was determined by a Gaussian fit  
performed on each of the 27 data sets collected over 
the four-month period of our tests. Two examples of
such fits are displayed in the top panel of  
Figure~\ref{fig:pix-4-gain-vs-time}.
As the data exhibit, the change of gain over 
the period of four months is small, on the order
of 5-10\%, and consistent with the expected 
aging of M16 PMTs~\cite{M16-1600}. The down
trend for the last two data acquisition periods,
persistent in several plots,  we attribute
to the change of temperature in our laboratory
which was about  $2-3^\circ$C lower 
than previously mentioned due to automatic 
air conditioning which was
beyond our control.

\begin{figure}[h]
\begin{center}
\includegraphics[width=.49\textwidth]
{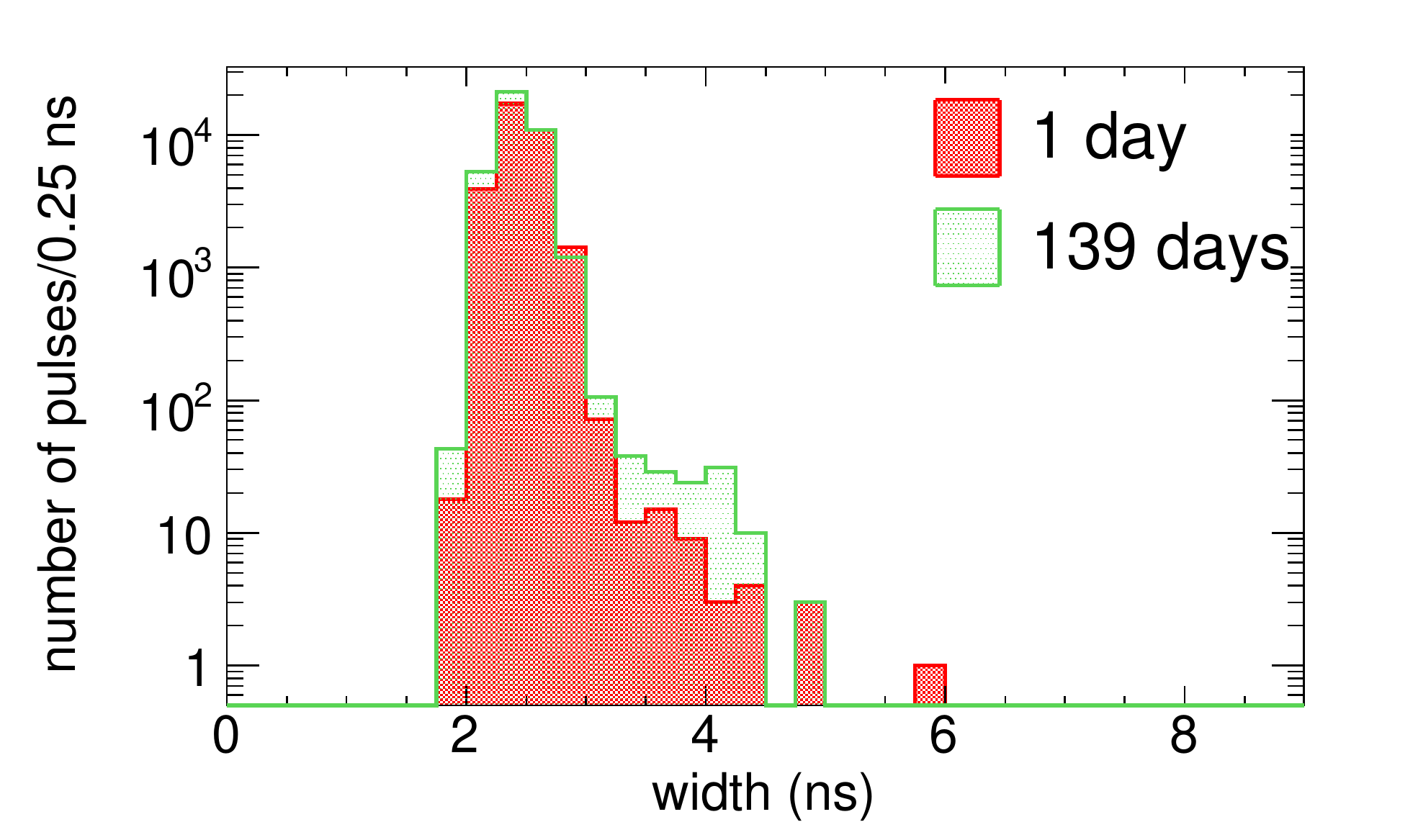}
\includegraphics[width=.49\textwidth]
{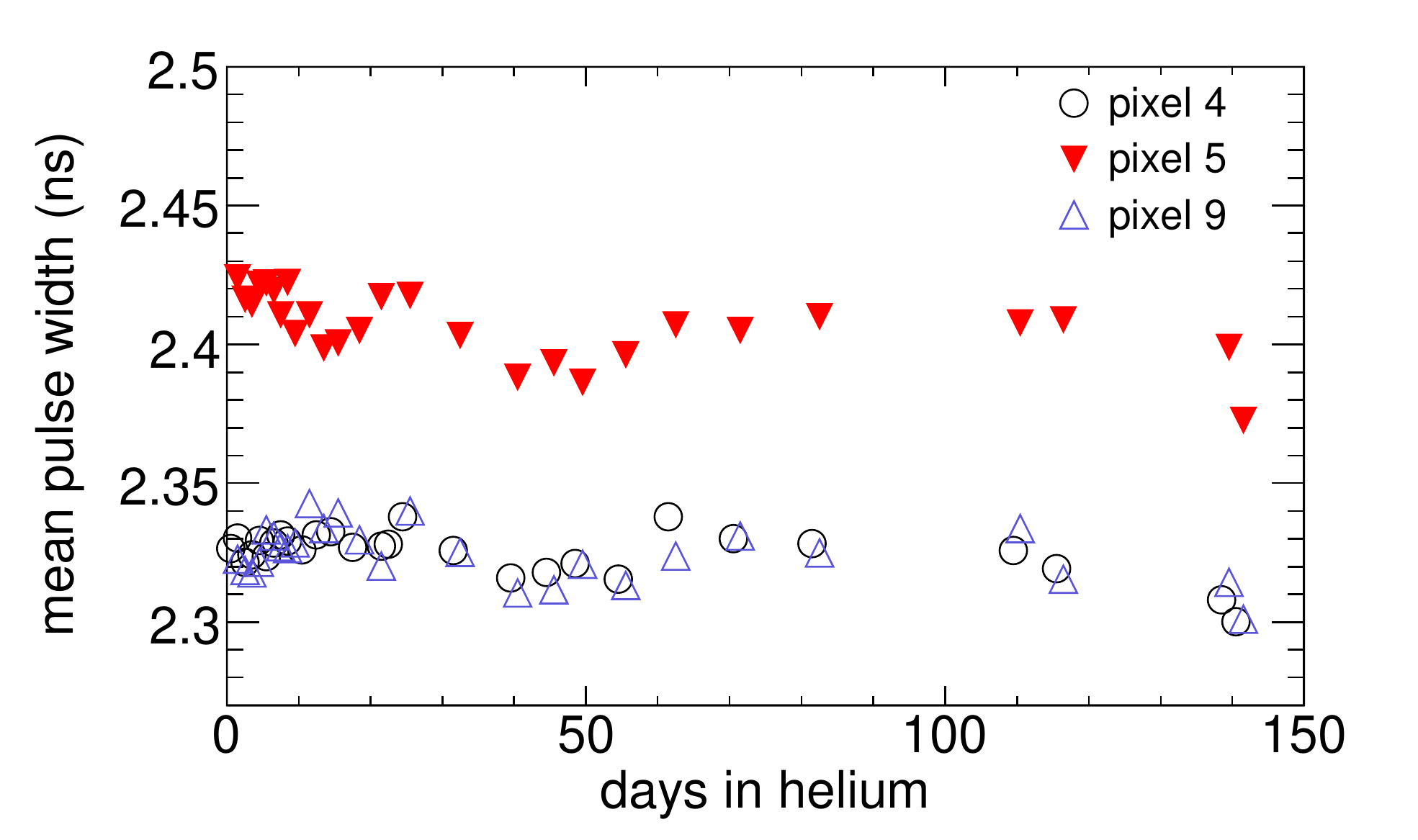}
\caption{ Width change of single photoelectron pulses.
Left: A distribution of the FWHM (in nanoseconds) of pulses 
for waveforms with a single pulse. The histogram is for 
the data taken at the beginning of our tests
and is representative of all the data throughout our tests.
Right: The average FWHM of pulses in waveforms with a single
pulse as function of the exposure time to helium. 
}
\label{fig:pix-4-width-vs-time}
\end{center}
\end{figure}

Figure~\ref{fig:pix-4-width-vs-time} shows  the widths 
of the single photoelectron pulses for the waveforms 
with a single pulse. We use the full-width-at-half-maximum 
(FWHM), expressed in nanoseconds, as a measure of the width 
of a pulse. The data show little change over the four-month
period of our tests. As indicated above, the last two points 
were taken at lower laboratory temperature.

\subsection{Gains and widths of after-pulses}

\begin{figure}[t]
\begin{center}
\includegraphics[width=.49\textwidth]
{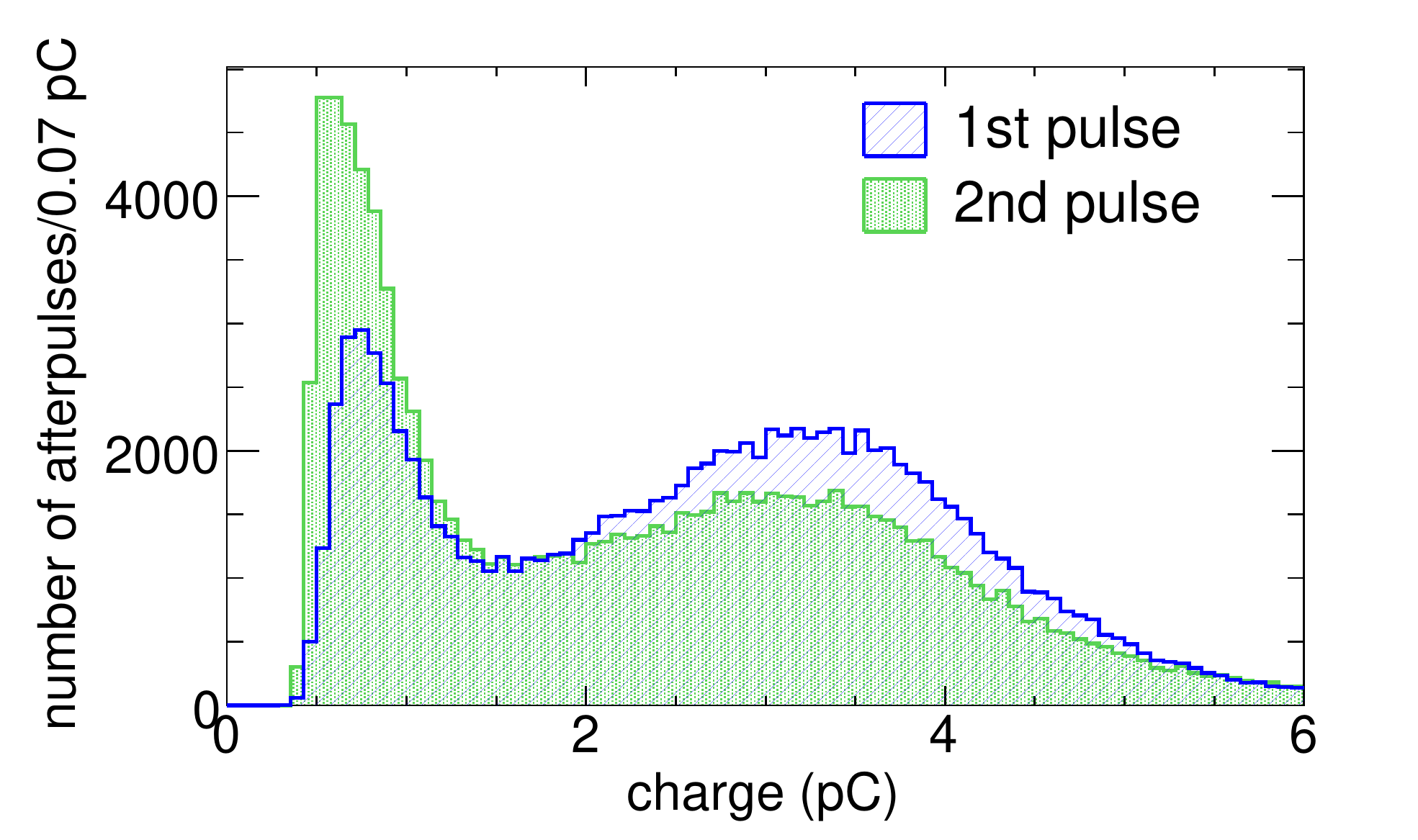}
\includegraphics[width=.49\textwidth]
{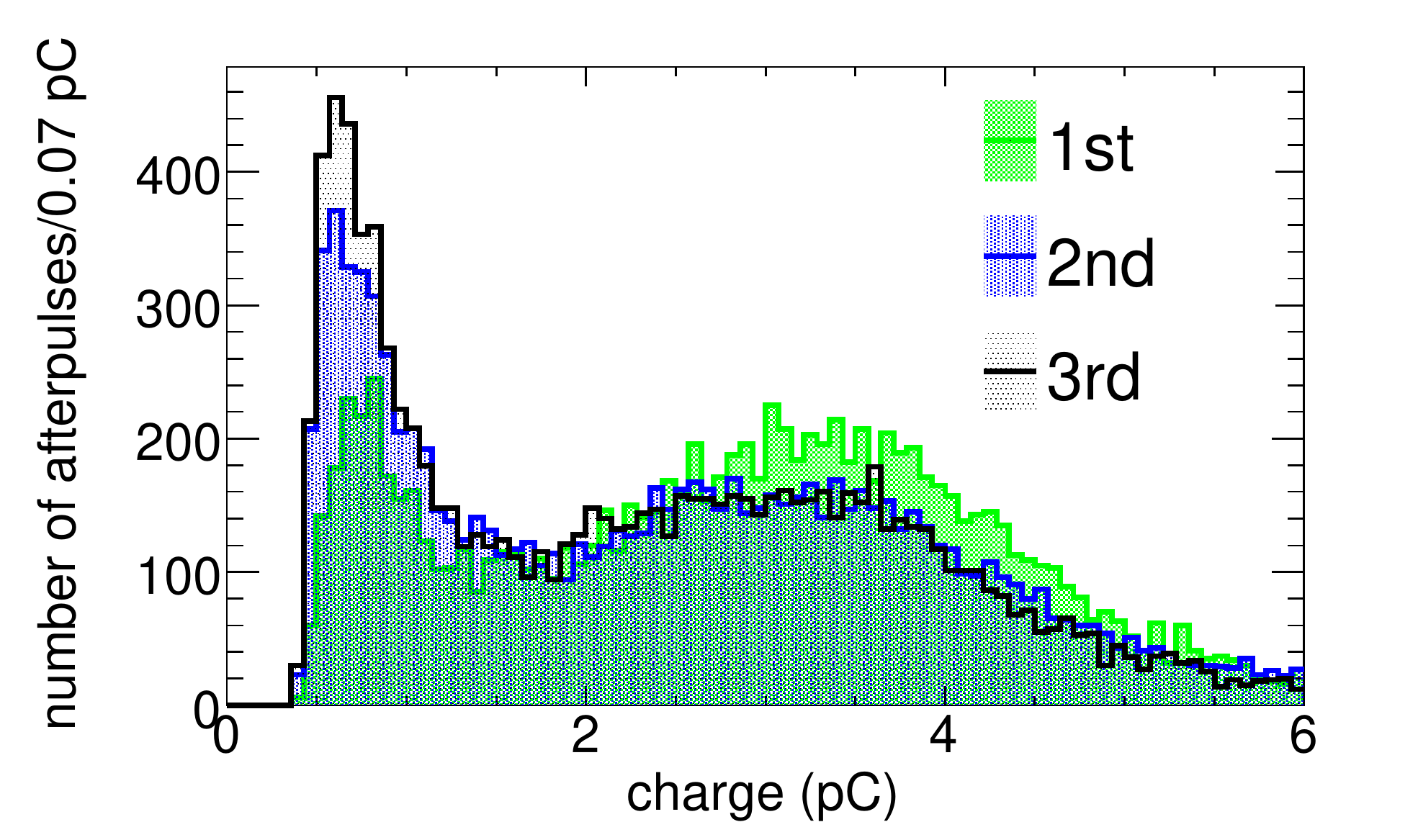}
\caption{Gains of after-pulses.
Left: Superposed distributions of charges of 
the first and the second pulse 
for the waveforms with two pulses.  
The first pulse has a peak at $3.21\pm0.01$~pC
and the second is at $3.08\pm0.01$~pC.
Right: Superposed distributions of charges 
of the first, the second, and 
the third pulse for the waveforms with three pulses.
The second  pulse has a peak at $3.04\pm0.03$~pC
and the third is at $3.02\pm0.05$~pC.
These distributions were obtained using all waveforms collected
over the entire four-month data-taking period.
}
\label{fig:pix-4-charge-of-1st-and-2nd}
\end{center}
\end{figure}

The after-pulses were defined as all pulses following the first
pulse within the \unit{400}{ns} time window of read out waveforms.
Our algorithm was able to clearly identify such waveforms. 
We first examined the gains of pulses for waveforms with exactly 
two pulses.
In the top panel of Figure~\ref{fig:pix-4-charge-of-1st-and-2nd} 
we compare distributions of the pulse charge for the first (in time)
and the second pulse within the \unit{400}{ns} time window.
We then compared gains of pulses for waveforms with exactly 
three pulses. The distributions of the pulse charges for 
the three pulses are displayed in the bottom panel of 
Figure~\ref{fig:pix-4-charge-of-1st-and-2nd}.

\begin{figure}[t]
\begin{center}
\includegraphics[width=.49\textwidth]
{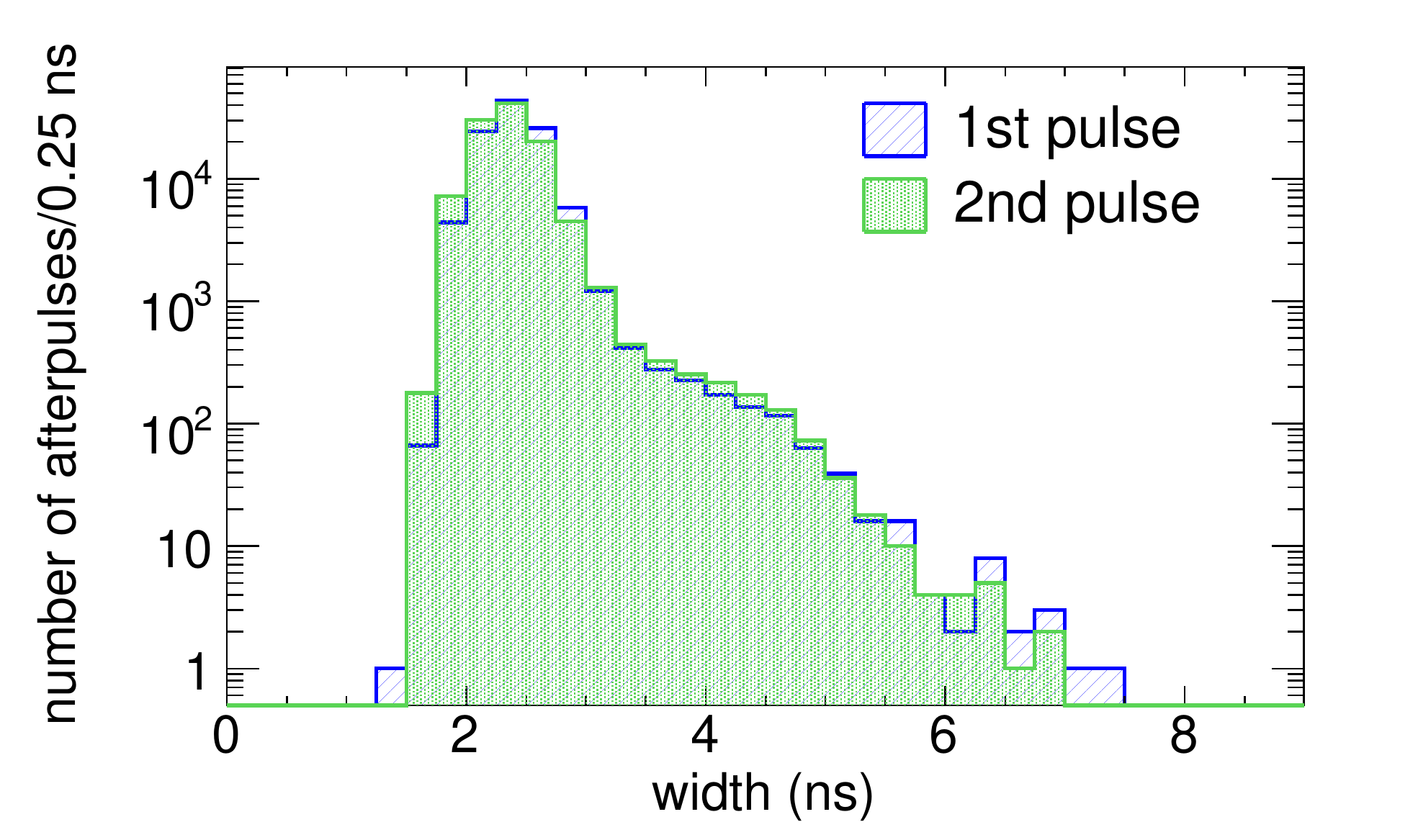}
\includegraphics[width=.49\textwidth]
{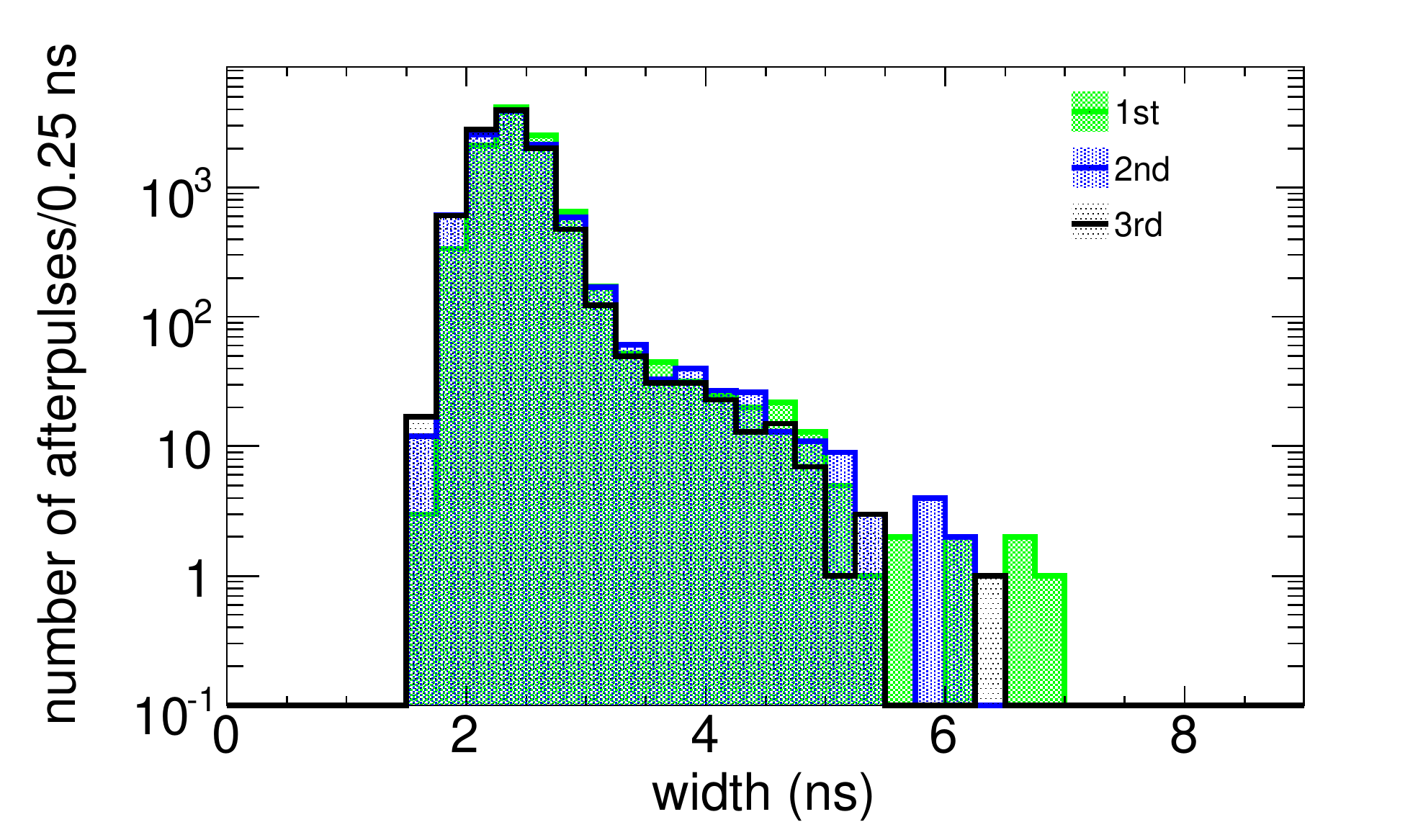}
\caption{Widths of after-pulses.
Left: Superposed distributions of the FWHM for the first and 
the second pulse for the waveforms with two pulses.
Right: Superposed distributions of the FWHM for the first, 
the second, and the third pulse for the waveforms with three pulses. 
These distributions were obtained using all waveforms collected
over the entire four-month data-taking period.
}
\label{fig:pix-4-width-of-1st-and-2nd}
\end{center}
\end{figure}

We note that the data clearly indicate that the gains of
all pulses  (i.e., the gain of the first pulse 
in a waveform, and the gains of the after-pulses), 
are very similar and do not change with the exposure time to helium.
Figure~\ref{fig:pix-4-width-of-1st-and-2nd} shows similar comparisons 
for the pulse widths for waveforms with after-pulses. 

We conclude that all pulses in the waveforms with multiple
pulses are similar, although the distributions show that 
the first pulse in such waveforms carries a little more charge 
than after-pulses, as shown in Figure~\ref{fig:pix-4-width-of-1st-and-2nd}. 
This may be a result of the complexity
of the ionization process of helium (or other residual gases)
which occasionally may produce two ionization electrons.
In such a rare event, an after-pulse would be preceded by 
two almost simultaneous electrons initiating a dynode
cascade identified as the first pulse.
 
The mean charge 
of all pulses is about \unit{3}{pC} (i.e., the gain of about 
$1.9\times 10^{7}$) and their average widths (FWHM) remained almost
constant during the four-month exposure to helium. The observed
variations with time are consistent with the expected aging 
of M16 PMTs observed previously~\cite{M16-1600}.

\subsection{Time structure of the M16 after-pulsing }

\begin{figure}[b]
\begin{center}
\includegraphics[width=.49\textwidth]
{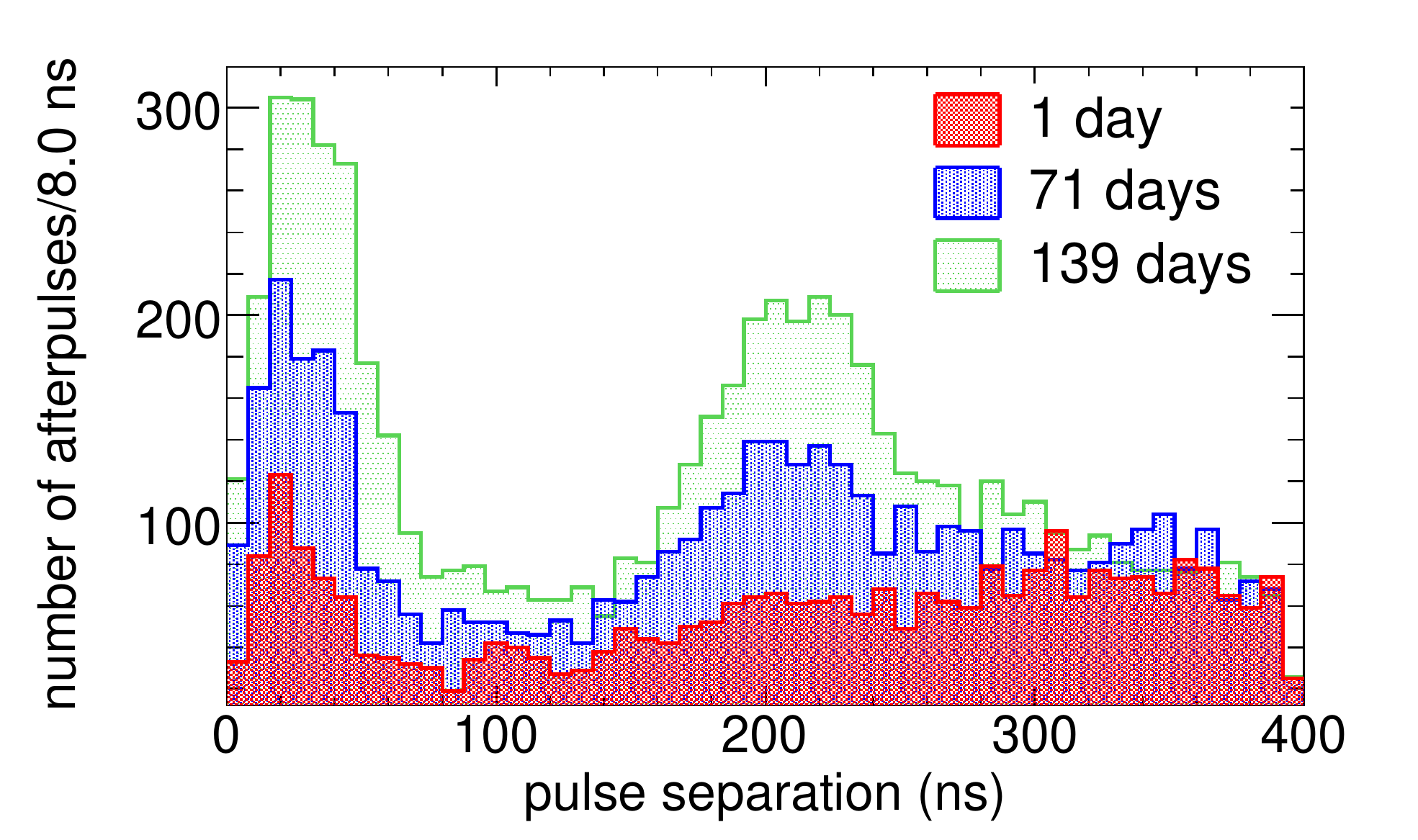}
\includegraphics[width=.49\textwidth]
{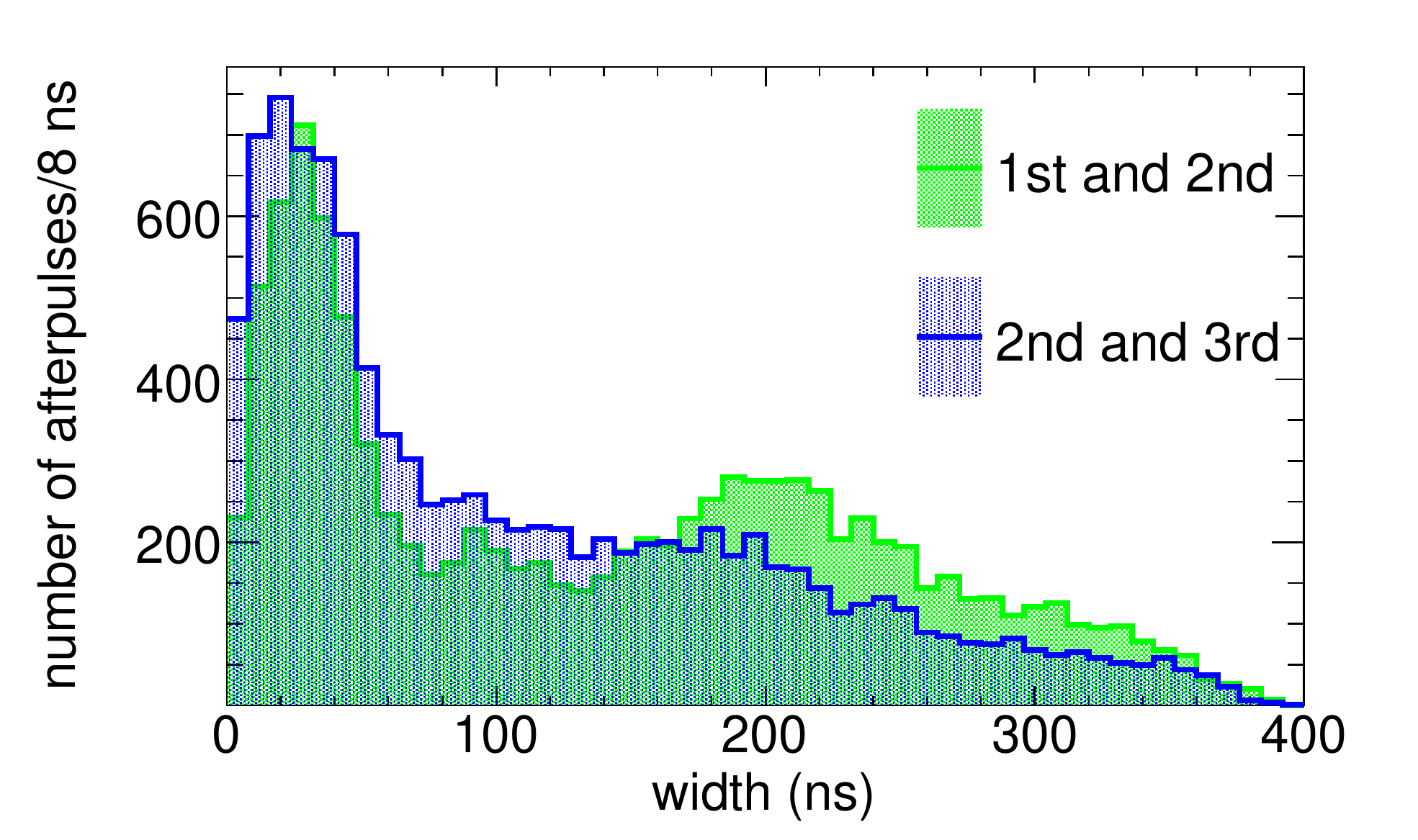}
\caption{Time separation between pulses.
Left: The time period separating two pulses in waveforms with two pulses. 
The three superposed distributions
were obtained using data taken about two months apart 
(i.e., at the beginning, in the middle, and at the end of our tests).
Right: The time separation between pulse 1 and 2, and between pulse
2 and 3, in waveforms with three pulses. 
These distributions were obtained using all waveforms collected
over the entire four-month data-taking period.
}
\label{fig:pix-4-time-separation}
\end{center}
\end{figure}

We then investigated the time structure of waveforms 
with after-pulses. In the top panel of  
Figure~\ref{fig:pix-4-time-separation} we display the time separation 
 between pulse 1 and 2 in waveforms with two pulses. 
The superposed distributions were obtained using three data sets 
taken two months apart  (i.e., at the beginning, in the middle, 
and at the end of our tests). 
In the bottom panel of  
Figure~\ref{fig:pix-4-time-separation} we compare
the time separation between pulse 1 and 2 with the 
time separation  between pulse 2 and 3 in waveforms 
with three pulses. These distributions
exhibit similar, althought not the same,
time characteristics.

The histograms in Figure~\ref{fig:pix-4-time-separation}  
show after-pulsing time structure 
with two distinct maxima. 
As our estimates in section 4 indicate,
the first peak, at about \unit{10-70}{ns},
is most likely due to after-pulsing caused by helium ionized between 
the photocathode and the first dynode. This is because
helium ions require about \unit{30}{ns} to accelerate across 
the distance between the first dynode and the photocathode.
The origin of the second broad peak at \unit{150-400}{ns} 
with a long right shoulder is more speculative.
We believe that it is due to after-pulses generated by helium 
ions produced ``deeper'' in the dynode channel
(i.e., ``below'' the first dynode) or in the peripheral 
electric field between dynodes. 
Some fraction of them may
be due to heavier ions. As Figure~\ref{fig:pix-4-time-separation} shows, 
all these features become more prominent with the increasing exposure 
time to helium (i.e., the higher helium poisoning of the PMT).

\begin{figure}[b]
\begin{center}
\includegraphics[width=.49\textwidth]
{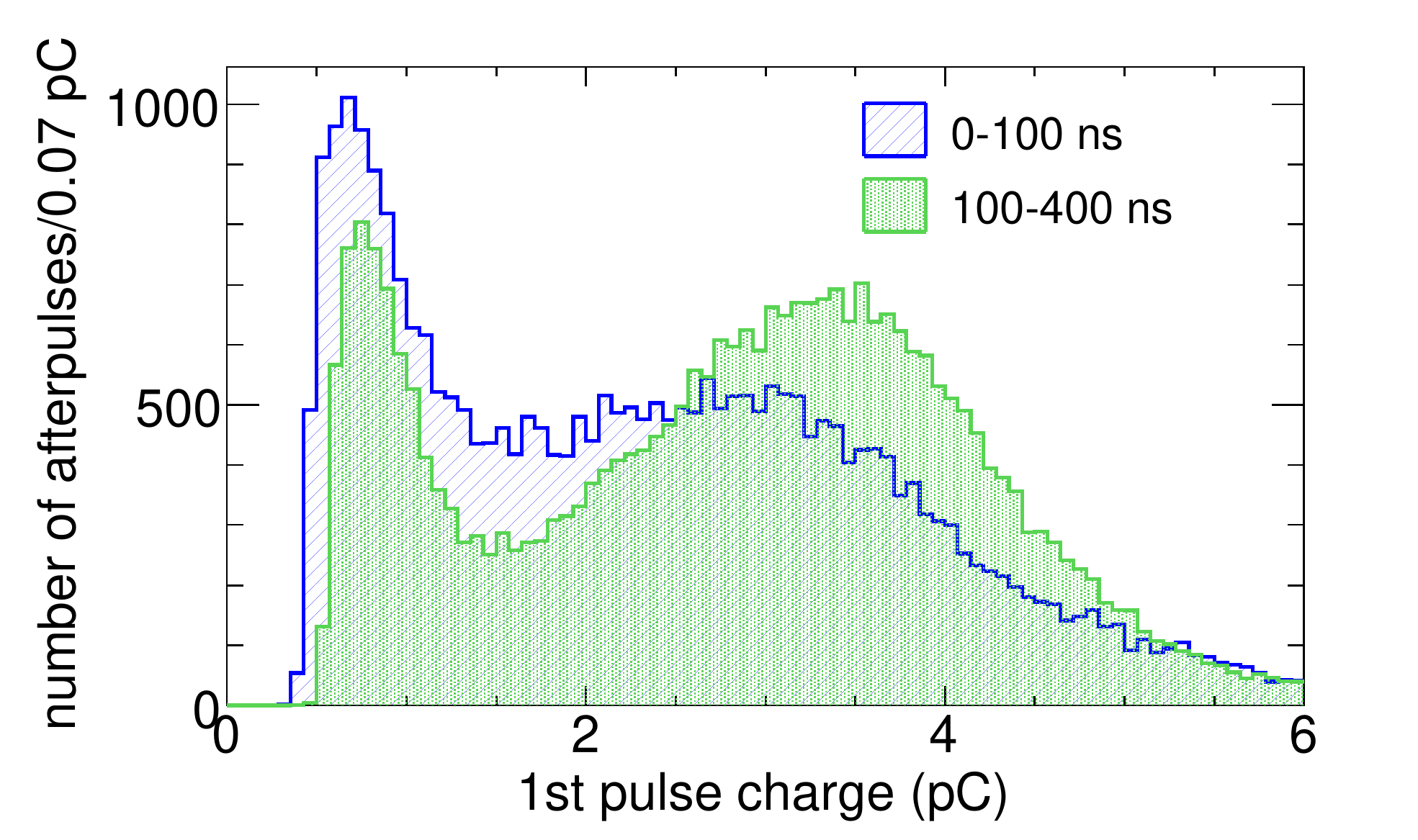}
\includegraphics[width=.49\textwidth]
{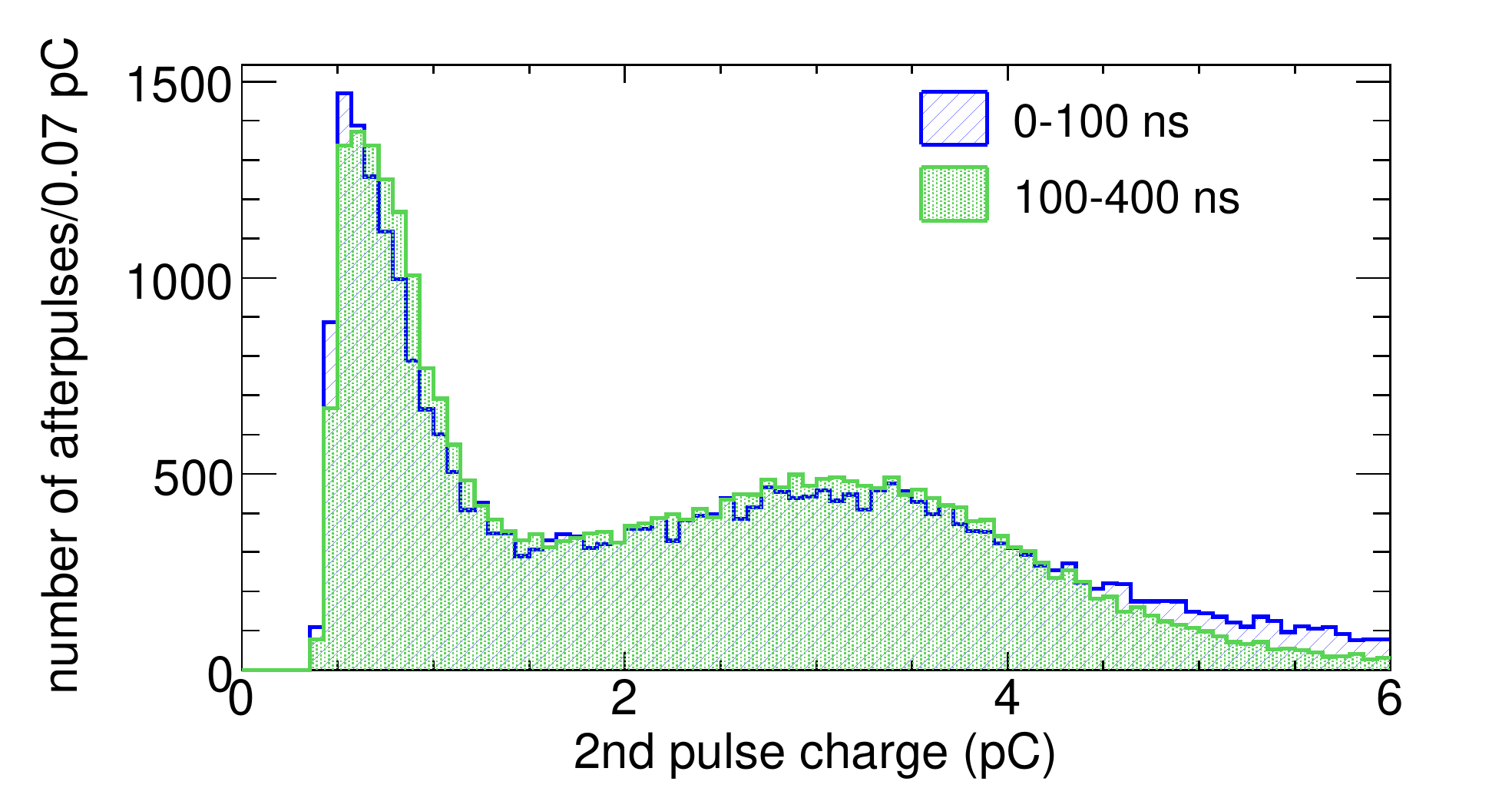}
\caption{Left: Superposed distributions of the charges in 
the first peak for waveforms in which the after-pulse
is separated from the first (trigger) peak by less
than \unit{100}{ns} or seprated by time between
100 to \unit{400}{ns}.
The first pulse in the  \unit{0-100}{ns} pulse category 
has a peak at $3.29\pm0.01$~pC
and  for the 100 to \unit{400}{ns} category the 
peak is at $2.72\pm0.04$~pC.
Right: Superposed distributions of the charge in
the second pulse for the same pulse categories. 
The second pulse in the  \unit{0-100}{ns} pulse category 
has a peak at $3.00\pm0.02$~pC
and  for the 100 to \unit{400}{ns} category the 
peak is at $3.06\pm0.01$~pC.
These distributions were obtained using all waveforms collected
over the entire four-month data-taking period.
}
\label{fig:pix-4-gain-various-times}
\end{center}
\end{figure}

The distributions of the charge carried by the first and the second
pulse for pulses separated by either less or more than \unit[100]{ns}, 
summarized in Figure~\ref{fig:pix-4-gain-various-times},
tend to support the idea of different location of the helium 
ionization.

\section{Phenomenological estimates of the helium poisoning of M16}

In this section we present simple calculations 
related to the observed after-pulsing due to helium poisoning.
Because the exact material properties and the internal 
dimensions of the M16 PMTs constitute proprietary 
information, we limit our considerations to 
``back of the envelope'' arguments.

We first estimate the helium permeation rate into the M16 PMT
in our setup. We then use the electron-helium ionization 
cross-section to find the probability for after-pulsing. 
Finally, we estimate the most probable time of after-pulses 
due to helium ionization.

Probability of after-pulsing depends on a mean-free-path 
for ionization by photoelectrons ejected from a photocathode, and
the distance and voltage between the photocathode and the first dynode. 
For simplicity, the first dynode and the photocathode are treated
as infinite conducting parallel plates. 

\subsection{Helium permeation in M16 PMTs}

In our calculations we neglect permeation of helium
through the metal walls of the M16. However, helium permeates          
easily through most glasses due to a small size 
and high mobility of helium atoms. In the following estimates
of the the helium permeation through the front window
of the M16 we also ignored the thin layer of 
photocathode~\cite{proprietary}.

The rate of permeation depends on the specific glass 
material, its thickness, the gradient of the helium partial 
pressure between the two sides of the glass barrier, and 
the ambient temperature, as given by a steady-state 
form of Fick's law~\cite{Fick-law}:
$$
{{dV} \over {dt}} = {{K \cdot A} \over {D}}\cdot \Delta P, 
$$
where $dV$ is a volume of helium at $STP$ which permeates
through the surface of area $A$ through a barrier of 
thickness $D$ in time $dt$. $\Delta P$ is the gradient of
the partial pressure of helium and
$K$ is the permeation constant which
depends on the barrier material.

The value of $K$ for borosilicate glass~\cite{M16-glass} 
can be found in two comprehensive compilations~\cite{shelby,altemose}:
$$
K = 4.8\times e^{-Q/RT_g}\left[ {\rm {{~cm^3} \cdot mm
\over { s \cdot cm^2  \cdot (cm ~Hg)}} }\right]
$$
where for the M16 glass the activation energy $Q=$~\unit[7,930]{cal/mol},
the gas temperature $T_g=$~\unit[294.15]{K}, and the gas constant 
$R=$~\unit[1.986]{cal/mol $\cdot$ K}.
These values give
$$
K = 6.1 \times 10^{-13} \left[ {\rm {{~cm^3} \cdot mm
\over { s \cdot cm^2  \cdot (cm ~Hg)}} }\right] 
$$
For the M16 PMT, $A \approx 5.76 ~{\rm cm^2}$, $D = 0.8 {\rm mm}$, and 
$\Delta P \approx 76 ~{\rm cm~Hg}$, hence
$
{{~dV} \over {~dt}} = 3.34 \times 10^{-10} {\rm cm^3 / s}. 
$
The volume of permeated helium after one day of exposure to 
pure helium gas at an atmospheric pressure ($\Delta P ~ \approx ~101~kPa$), is:
$
V_{He}({\rm 1 ~day}) =3.34 \times 10^{-10} {\rm ~[cm^3/s] 
\times 24 \times 3600 ~{\rm [s]} } \approx 2.88 \times 10^{-5} ~{\rm cm^3}.
$
We estimate the gas volume of the M16 tube to be about \unit[10]{cm$^3$}, thus 
after one day, the partial pressure of helium inside the tube is (at STP):
$$
P_{He}(1~day) = {V_{He}(1~day)  \over V_{tube}} \times 101 {\rm ~[kPa]} 
\approx 0.29 {\rm ~Pa}.
$$
This corresponds to $N_{He}({\rm 1~day}) \approx 7.75 \times 10^{14}$ helium atoms 
penetrating through the glass window of the M16 PMT in one day. Thus, we expect
that at the end of our tests, which lasted four months, the number of
helium atoms which permeated into the tube to be 
$N_{He}{\rm (120~days)} \approx 9.3\times 10^{16}$,
which corresponds to the partial helium
pressure inside the M16 PMT of $P_{He}{\rm (120~days)} \approx 35 {\rm ~Pa}$.
We will use these results to estimate the effective 
mean-free-path for helium ionization and the probability of 
the after-pulsing.

\subsection{Ionization of helium atoms in M16 PMTs}

\begin{figure}[b]
\centerline
{\includegraphics[width=.79\textwidth]{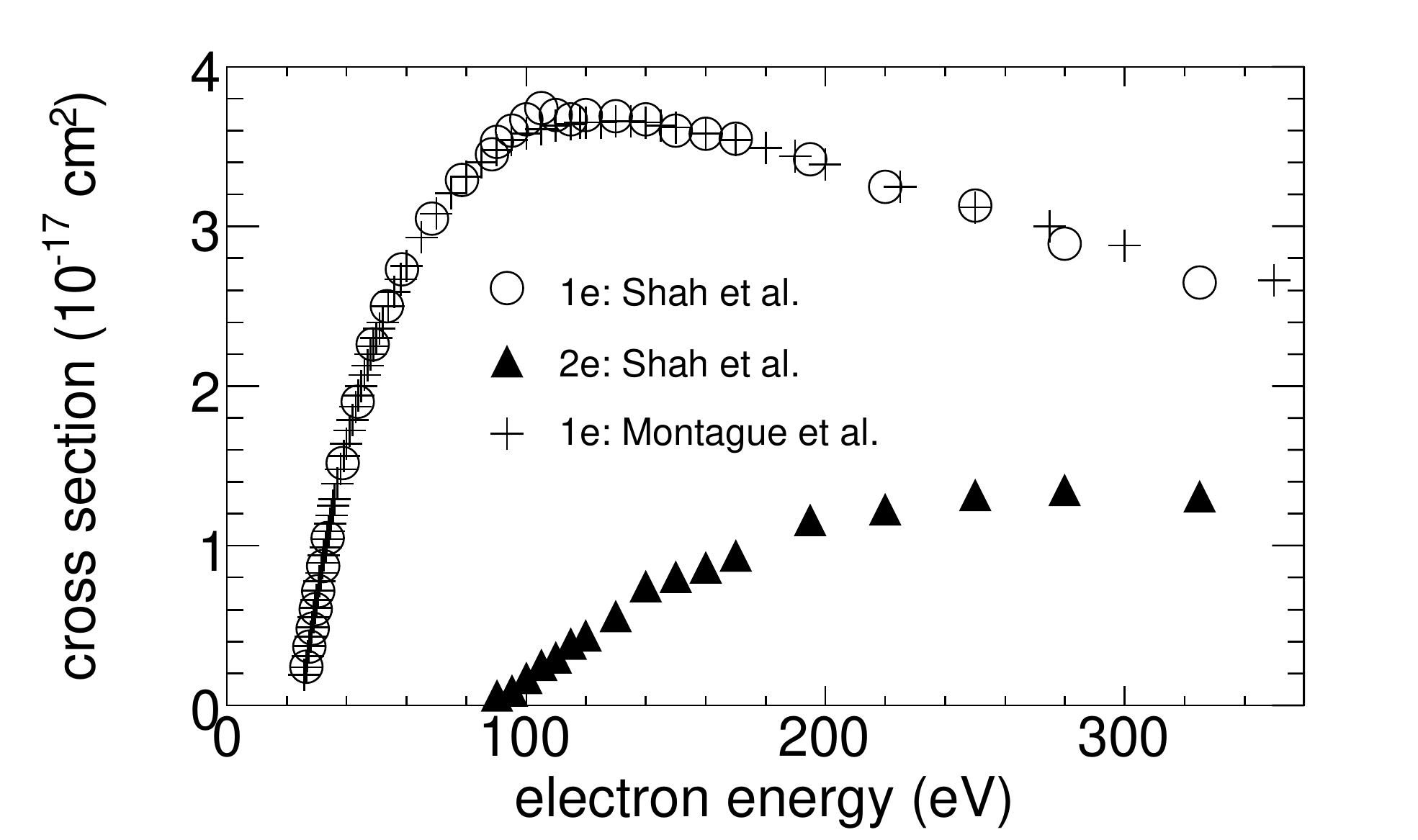}}
\caption{ The total ionization cross-section of a helium 
atom by electrons as a function of the electron energy.
The plot is reproduced from ~\cite{Triangles} (triangles) and from
~\cite{Circles} (circles). The two distinct sets of data 
are for a single-electron ionization (higher set of points) and 
a double-electron ionization (lower set of points). } 
\label{fig:he-ion}
\end{figure}

The total ionization cross-section of helium atoms depends on 
the energy of the impacting electron and is well measured 
for a wide range of energies, as shown in Figure~\ref{fig:he-ion}
which was reproduced from references~\cite{Triangles} and
\cite{Circles}. The cross-section for
a single-electron ionization is essentially zero below
\unit[25]{eV} and then grows rapidly reaching a maximum at
about \unit[120]{eV}; it decreases slowly as the energy of
an impacting electron increases. The ionization
probability is most significant at an energy of about 
\unit[80-120]{eV}. 

Photoelectrons are accelerated inside 
the PMT in the electric field between the photocathode 
and subsequent dynodes. In our tests we used a standard 
MINOS M16 base with the 12-stage voltage divider which
has the photocathode to anode voltage ratios of
2.4-2.4-2.4-1.0-1.0-1.0-1.0-1.0-1.0-1.0-1.2-2.4.
The nominal high voltage values in MINOS were set
to achieve a gain of about $1\times10^6$, which can
be reached for most PMTs at \unit{700-900}{V}~\cite{M16-1600}. 
In our helium poisoning tests, however, we used the
maximum recommended high voltage of \unit[1000]{V} to
maximize the PMT gain, the photoelectron collection efficiency,
and the cross-section for helium ionization.
For such a setting the voltage difference between 
the photocathode and the first dynode, $\Delta V_{1}$ is 
$
\Delta V_{1} = {R_{1} \over R_{total}} \times 1000 ~{\rm [V]} = 
{2.4 \over 17.8} \times 1000 ~{\rm [V]} = 135 ~{\rm V}.
$
We considered changing the voltage divider ratios  
but decided against it since a modified base would significantly 
alter operational characteristics of the tube (otherwise
extensively studied earlier~\cite{M16-1600}).
A photoelectron is accelerated in the electric field given by
$E = \Delta V_{1} / d$, where $d \approx 1.5 ~{\rm mm}$
is the distance from the photocathode to the first dynode.
\footnote{The exact value of this distance is not known to us
since this is proprietary information.
Here, we provide our best estimate.} 
 
The cross-section for ionization grows as the photoelectron gains
energy as it approaches the first dynode.
The mean-free-path for helium ionization, $\lambda_{He}^{ion}$, 
is given by:
$$
\lambda_{He}^{ion} = {1 \over {n_{He} \cdot \sigma_{He}^{ion}}} = 
{1 \over {N_{He}/V_{tube} \cdot \sigma_{He}^{ion}}},
$$
where $n_{He}$ is the density of helium atoms in the tube,
and $\sigma_{He}^{ion}$ is the cross-section for helium ionization.
For our estimates we take somewhat arbitrarily 
$ \sigma_{He}^{ion} = 3\times 10^{-17}~{\rm cm^2} $ 
and the estimated earlier value for $N_{He}$, after one day 
in the helium atmosphere we calculate
$ \lambda_{He}^{ion}({\rm 1~day}) \approx 430 ~{\rm cm} $, 
and after the four-month exposure to helium
$ \lambda_{He}^{ion}({\rm 120~days}) \approx 3.6 ~{\rm cm}$.

\subsection{Estimate of the probability of M6 after-pulsing}

Most after-pulsing is caused by positive helium ions
drifting to the photocathode, scattering off it, and causing 
additional photoelectron emission. Thus, the density
of helium ions is closely coupled to the probability of
after-pulsing. The ionization probability of helium atoms 
inside M16 is proportional to the ratio 
of the electron path-length between the photocathode and 
the first dynode to the mean-free-path for helium ionization,
$ \lambda_{He}^{ion}$. The sharp rise of
$\sigma_{He}^{ion}$ with energy means that the ionization process 
is more likely to occur {\it near} the first dynode, where 
photoelectrons are sufficiently energetic.
We can {\it crudely} estimate the ionization probability 
by assuming the {\it effective } path-length to be about 
$ {1\over 3} d = 0.5 ${mm}. 
Thus, after one day in helium, the ionization probability, 
$P_{{\rm 1~day}}$,  is given by 
$P_{{\rm 1~day}} = 0.05~{\rm cm}/430~{\rm cm} 
\approx 1.2\times 10^{-4} = 0.012\%$,
while for 120 days 
$P_{{\rm 120~days}} = 0.05~{\rm mm}/3.6~{\rm cm} 
\approx 0.014 = 1.4\%$.
In our simple model, we equate this ionization probability 
with the probability of after-pulsing.

\subsection{Timing of after-pulses}\label{section-time}

As discussed above, the ionization of helium occurs when a photoelectron 
has sufficiently high energy, which for the M16 PMT  occurs near the first 
dynode.  The helium ion is then accelerated to the photocathode and may become 
the source of a secondary electron. Its travel time determines the time 
separation between the initial (helium ionizing) photoelectron
and an electron ejected by an impact of a helium ion with the photocathode.
This time can be roughly estimated by the time-of-flight of a helium ion
along a straight line through the distance between the photocathode
and the first dynode: 
$$ 
t = \sqrt{ {2d \over a} }= \sqrt{ {2 d m_{He} } \over {{\Delta V_1 \over d} e}} 
= d \sqrt{ {2 m_{He} } \over {{\Delta V_1  e}}} = 27 {\rm~ns},
$$
where $d = 1.5~{\rm mm}$, $\Delta V_1 = 135~{\rm V}$,  and $a$ are a distance, 
voltage, and an acceleration of a helium ion between the photocathode and 
the first dynode, $m_{He}$ is its mass, and $e$ is a unit of an electric charge. 
In reality, the timing of helium ions will depend on the actual shape and
strength of the electric field, and  the place where the helium atom
was ionized, so the time separation of the 
after-pulsing would be smeared around this average value.

\section{Discussion}

The presented estimate of the probability and 
timing of after-pulses roughly agrees with main features 
of the recorded data. The probability of after-pulsing 
increased with the exposure time to helium
by about 1 to 4\%, depending on the pixel.  A significant
fraction of after-pulses is attributed to the initial residual
contamination of the M16 vacuum.

The time separation of observed after-pulses is consistent with
the process of helium ionization by photoelectrons. The ionization 
can occur either between the first dynode and the photocathode 
or deeper in the dynode channel structure of the M16.
Only after the photoelectrons gain sufficient energy and are close to 
the first dynode does the ionization cross-section become significant. 
In the M16 PMT this effectively reduces the mean-free-path for 
an electron to ionize helium atoms.

We believe that the observed resilience to helium permeation 
and the low probability of after-pulsing are due to the M16 
compact dynode structure housed in a metal casing which 
operates at a negative potential of the photocathode. 
The electron energies achieved through the acceleration 
in the electric fields inside an M16 PMT are relatively
small. This is because they depend on the voltage differences 
between dynodes which for the specific M16 voltage divider
used by MINOS and the maximal operating HV of \unit{1000}{V}
range from \unit{56}{V} to \unit[135]{V}. Short
distances that electrons travel inside the tube and low voltage
differences give the low probability of after-pulses as compared
to larger photomultipliers with higher accelerating potentials, 
as for example studied in reference~\cite{other-studies}.

We studied after-pulses related 
only to dark noise single photoelectron pulses. Since signal 
light levels in MINOS are of the order of several to few tens of 
photoelectrons, our results are relevant to the MINOS detectors
(although MINOS PMTs operated at a lower gain)
but may not be for applications with significantly larger signals.

\section{Conclusions}

We have studied the effects of helium poisoning 
of R5900-00-M16 PMTs which were employed in the MINOS 
Far detector. A photomultiplier tube was immersed in
pure helium atmosphere for about four months
and its response was monitored using a digital oscilloscope.
During this time, the PMT showed a slow increase 
in the probability of after-pulsing but its main performance 
characteristics degraded only slightly. The observed decrease
of gain by a few percent is consistent with a typical aging 
of such PMTs~\cite{M16-1600}. 

In our tests, the primary pulses were caused by dark noise
single photoelectrons. We observed that the probability of 
after-pulsing grew linearly with the PMT exposure time to
helium, and that the primary pulses and the associated  
after-pulses have similar gains and widths.  We conclude that 
our data indicate that this type of PMTs exhibits 
resilient response to helium poisoning.

We thank Yuji Yoshizawa and Wayne Stehle of Hamamatsu 
Photonics for providing PMTs for our tests and 
furnishing additional technical information and assistance.  
This work was supported in part by 
DOE grant DE-FG03-93ER40757.

%
%
%


\begin{thebibliography}{999}

\bibitem{MINOS_Proposal} 
P875: A Long-baseline Neutrino Oscillation Experiment at
Fermilab, The MINOS Collaboration, Fermilab, February, 1995.

\bibitem{MINOS_TDR} 
The {\sc MINOS} Detectors, Technical Design Report, 
Fermilab, NuMI-L-337, Fermilab, October, 1998.

\bibitem{Lang:2001rw}
K.~Lang  [for MINOS Collaboration],
  Nucl.\ Instrum.\ Meth.\ A {\bf 461}, 290 (2001).
  
\bibitem{M16-1600}
  K.~Lang {\it et al.},
  Nucl.\ Instrum.\ Meth.\ A {\bf 545}, 852 (2005).

\bibitem{1st-MINOS-PRL}
First beam paper: D.~G.~Michael {\it et al.} [MINOS Collaboration],
   Phys.\ Rev.\ Lett.\  {\bf 97}, 191801 (2006);
  the most recent paper:
  P.~Adamson {\it et al.} [MINOS+ Collaboration],
  Phys.\ Rev.\ Lett.\  {\bf 122}, no. 9, 091803 (2019).

\bibitem{wls-fiber}
  S.~Avvakumov {\it et al.},
   Nucl.\ Instrum.\ Meth.\ A {\bf 545}, 145 (2005).

\bibitem{CDMS-2} 
The {\sc CDMS-II} (Cryogenic Dark Matter Search) Experiment
is located in the Soudan-2 cavern in the Soudan Underground
Laboratory in Minnesota. The cavern is neighboring the new 
MINOS cavern also excavated \unit{705}{m} underground.

\bibitem{LabView}
LabView is a product of National Instruments Corporation, 
11500 N Mopac Expwy, Austin, TX 78759-3504, USA.

\bibitem{Hamamatsu-private}
Private communication from Hamamatsu Photonics.

\bibitem{LeCroy}
We used a digital oscilloscope model 584L  made by 
LeCroy Corporation, 700 Chestnut Ridge Rd., 
Chestnut Ridge, NY 10977-6499, USA.

\bibitem{proprietary} 
The thickness and exact composition of the photocathode
constitutes proprietary information. We speculate that
the photocathode is a bialkali layer of thickness of
2,000 to \unit{10,000}{nm}
(i.e., a few hundred of atomic layers).

\bibitem{Fick-law}
R.~M.~Barrer, {\it Diffusion in and through Solids},
Cambridge University Press, New York, 1941.

\bibitem{M16-glass}
According to Hamamatsu Photonics, the composition 
of the M16 borosilicate glass front window is 
very close to Glass Designation 7056 with 
$Si0_2(68.0$\%), 
$Al_2O_3(3.0$\%),
$B_2O_3(18.0$\%),
$LiO_2(1.0$\%),
$Na_2O(1.0$\%),
$BaO(0.0$\%),
$K_2O(9.0$\%) 
which we used to find the value
of the permeation constant $K$ 
in reference ~\cite{shelby}.

\bibitem{shelby}
J.~E.~Shelby, {\it Handbook of Gas Diffusion in Solids
and Melts''}, ASM International, Materials Park, OH, 1996
(ISBN: 0-87170-566-4);

\bibitem{altemose}
V.~O.~Altemose, J.\ Appl.\ Phys.\ {\bf 32}(7), 1309 (1961).
  
\bibitem{Triangles}
M.~B.~Shah, D.~S.~Elliot, P.~McCallion and H.~B.~Gilbody,
J.\ Phys.\ B.\ At.\ Mol.\ Opt.\ Phys.\ {\bf 21}, 2751 (1988).
                                                                                   
\bibitem{Circles}
R.~G.~Montague, M.~F.~A.~Harrison and A.~C.~H.~Smith,
J.\ Phys.\ B.\ At.\ Mol.\ Opt.\ Phys.\ {\bf 17}, 3295 (1984).
                                                                                   

\bibitem{other-studies}
J.~R.~Incandela {\it et al.}, \ Nucl. \ Instrum. \ Meth. 
{\bf A269}, 237 (1988).

\end{thebibliography}
\end{document}